\documentclass[11pt]{article}

\usepackage{eepic,epic}

\usepackage{mathptmx}
\usepackage{amssymb}
\usepackage{amsmath}
\usepackage{pifont}

\usepackage[english,vlined,boxed]{algorithm2e}



\newcommand{\OMIT}[1]{} %

\newenvironment{proofs}{\noindent{\bf Proof.}\hspace*{1em}}{\literalqed\bigskip}
\def\literalqed{{\ \nolinebreak\hfill\mbox{\qedblob\quad}}}

\usepackage{ifthen}
\usepackage{nicefrac}
\newcommand{\hugeDebug}{false}
\ifthenelse{\equal{\hugeDebug}{false}}{
\newcommand{\normalspacing}{\singlespacing}
}
{
\newcommand{\normalspacing}{\niceninespacing}
}
\ifthenelse{\equal{\hugeDebug}{false}}{
\setlength{\oddsidemargin}{0.25in}
\setlength{\evensidemargin}{\oddsidemargin}
\setlength{\textwidth}{6in}
\setlength{\textheight}{8.08in}
\setlength{\topmargin}{-0.0in}
}
{
\setlength{\oddsidemargin}{-0.4in}
\setlength{\evensidemargin}{\oddsidemargin}
\setlength{\textwidth}{5.3in}
}

\ifthenelse{\equal{\hugeDebug}{false}}{
\pagestyle{plain}
}
{
\pagestyle{headings}
}

\makeatletter
\newcommand{\singlespacing}{\let\CS=
\@currsize\renewcommand{\baselinestretch}{1}\tiny\CS}
\newcommand{\singlespacingplus}{\let\CS=
\@currsize\renewcommand{\baselinestretch}{1.25}\tiny\CS}
\newcommand{\doublespacing}{\let\CS=
\@currsize\renewcommand{\baselinestretch}{1.75}\tiny\CS}
\newcommand{\extradoublespacing}{\let\CS=
\@currsize\renewcommand{\baselinestretch}{1.9}\tiny\CS}
\newcommand{\draftspacing}{\let\CS=
\@currsize\renewcommand{\baselinestretch}{2.0}\tiny\CS}
\newcommand{\hugedraftspacing}{\let\CS=
\@currsize\renewcommand{\baselinestretch}{2.4}\tiny\CS}
\newcommand{\niceonespacing}{\let\CS=\@currsize\renewcommand{\baselinestretch}{1.1}\tiny\CS}
\newcommand{\nicetwospacing}{\let\CS=\@currsize\renewcommand{\baselinestretch}{1.2}\tiny\CS}
\newcommand{\nicethreespacing}{\let\CS=\@currsize\renewcommand{\baselinestretch}{1.3}\tiny\CS}
\newcommand{\singlespacingplusplus}{\let\CS=\@currsize\renewcommand{\baselinestretch}{1.35}\tiny\CS}
\newcommand{\nicefourspacing}{\let\CS=\@currsize\renewcommand{\baselinestretch}{1.4}\tiny\CS}
\newcommand{\nicefivespacing}{\let\CS=\@currsize\renewcommand{\baselinestretch}{1.5}\tiny\CS}
\newcommand{\nicesixspacing}{\let\CS=\@currsize\renewcommand{\baselinestretch}{1.6}\tiny\CS}
\newcommand{\nicesevenspacing}{\let\CS=\@currsize\renewcommand{\baselinestretch}{1.7}\tiny\CS}
\newcommand{\niceeightspacing}{\let\CS=\@currsize\renewcommand{\baselinestretch}{1.8}\tiny\CS}
\newcommand{\niceninespacing}{\let\CS=\@currsize\renewcommand{\baselinestretch}{1.9}\tiny\CS}
\makeatother
\normalspacing

\makeatletter \@beginparpenalty=10000 \makeatother

\mathcode`\0="0030      %
\mathcode`\1="0031
\mathcode`\2="0032
\mathcode`\3="0033
\mathcode`\4="0034
\mathcode`\5="0035
\mathcode`\6="0036
\mathcode`\7="0037
\mathcode`\8="0038
\mathcode`\9="0039

\newcount\hour  \newcount\minutes  \hour=\time  \divide\hour by 60
\minutes=\hour  \multiply\minutes by -60  \advance\minutes by \time
\def\mmmddyyyy{\ifcase\month\or Jan\or Feb\or Mar\or Apr\or May\or Jun\or Jul\or
  Aug\or Sep\or Oct\or Nov\or Dec\fi \space\number\day, \number\year}
\def\hhmm{\ifnum\hour<10 0\fi\number\hour :%
  \ifnum\minutes<10 0\fi\number\minutes}
\newcommand\qedblob{\mbox{\ding{113}}}

\newtheorem{theorem}{Theorem}[section]

\newtheorem{definition}[theorem]{Definition}
\newtheorem{lemma}[theorem]{Lemma}

\newtheorem{proposition}[theorem]{Proposition}

\newtheorem{remark}[theorem]{Remark}

\newcounter{alg}

\newenvironment{fall}{\begin{list}
   {{\bf Case~\arabic{alg}:}}
   {\usecounter{alg}}}{\end{list}}

\newenvironment{unterfall}{\begin{list}
   {{\bf Case~\arabic{alg}.\arabic{subalg}:}}
   {\usecounter{subalg}}}{\end{list}}
\newcounter{subalg}

\title{Degrees of Guaranteed Envy-Freeness in Finite Bounded
Cake-Cutting Protocols\thanks{Supported
in part by the DFG under grants RO~\mbox{1202/12-1}
(within the European Science Foundation's EUROCORES program LogICCC:
``Computational Foundations of Social Choice'') and RO~\mbox{1202/11-1}
and by the Alexander von Humboldt Foundation's TransCoop program.
A preliminary version of this paper is to appear in the proceedings of
the 5th Workshop on Internet \& Network Economics (WINE-2009),
December 2009.  Work done in part while the second author was visiting
the University of Rochester.} 
}

\author{Claudia Lindner \ and \ J{\"o}rg Rothe \\
        Institut f\"ur Informatik \\
        Heinrich-Heine-Universit{\"a}t D{\"u}sseldorf  \\
        40225 D\"usseldorf, Germany
       }

\date{October 19, 2009}

\begin{document}

\maketitle

\begin{abstract}
Cake-cutting protocols aim at dividing a ``cake'' (i.e., a divisible resource)
and assigning the
resulting portions to several players in a way that each of the players
feels to have received a ``fair'' amount of the cake.  An important
notion of fairness is envy-freeness: No player wishes to switch the
portion of the cake received with another player's portion.  Despite
intense efforts in the past, it is still an open question whether
there is a \emph{finite bounded} envy-free cake-cutting protocol for
an arbitrary number of players, and even for four players.

We introduce the notion of degree of guaranteed
envy-freeness (DGEF) as a measure of how good a
cake-cutting protocol can approximate the ideal of envy-freeness while
keeping the protocol finite bounded (trading being disregarded).  
We propose a new finite bounded
proportional protocol for any number $n \geq 3$ of players, and show that
this protocol has a DGEF of $1 + \left\lceil \nicefrac{n^2}{2}\right\rceil$.
This is the currently best DGEF among known finite bounded cake-cutting 
protocols for an arbitrary number of players.  We will make the case
that improving the DGEF even further is a tough challenge, and
determine, for comparison, the DGEF of selected known
finite bounded cake-cutting protocols.
\end{abstract}

\section{Introduction}
\label{sec:introduction}

Fair allocation of goods or resources among various agents is a 
central task in multiagent systems and other fields.  The specific
setting where just one divisible resource is to be divided fairly is
commonly referred to as cake-cutting, and agents are called
players in this setting.
Research in the area of cake-cutting
started off in the 1940s with the pioneering work of
Steinhaus~\cite{ste:j:steinhaus} who, 
to the best of our knowledge, was the first
to introduce the problem of fair division.
Dividing a good
(or a resource) fairly among
several players such that each of them is satisfied with the portion
received is of central importance in many fields.
In the last 60 years this research area has
developed
vividly, spreading out into various directions
and with applications in areas as diverse
as economics, mathematics, computer science, and
psychology. While some lines of this research
seek to find
reasonable interpretations of what
``fairness'' really stands for and how to
measure it~\cite{fol:j:resource-allocation,fel-kir:j:fairness-and-envy,cha:j:measure-envy}, others are
concerned more with proofs of existence or impossibility theorems
regarding fair division
\cite{var:j:equity-envy-efficiency, wel:j:fair-division,aki:j:vilfredo-pareto}, 
or with the development of new cake-cutting
procedures~\cite{bra-tay:j:protocol,str:j:cut-cake,rob-web:b:cake-cutting,bar-bra:j:minimal-cuts}
and, relatedly, with the analysis of their complexity
relative to both upper and lower bounds
\cite{mag-bus-kri:c:cake-cutting,woe-sga:j:complexity-cake-cutting,pro:c:cake-cutting}.
Since cake-cutting procedures involve several parties (namely, the
players), they are also referred to as ``protocols.''

Cake-cutting protocols
aim at achieving a \emph{fair} division
of an infinitely divisible resource among $n$ players,
who each may have different preferences for
(and thus different valuations of) different parts of the
resource. In this paper, we
focus on
(a notion of) fairness in
finite bounded cake-cutting protocols.
Many cake-cutting protocols are known today, both finite and
continuous ones. While a \emph{finite} protocol
always provides a solution after
only a finite number of decisions have been made, a \emph{continuous}
protocol could potentially run forever.
Among finite protocols, one can further
distinguish between bounded and unbounded ones.  A finite \emph{bounded}
cake-cutting protocol is present if we know in advance that a
certain number of steps (that may depend on the number of players)
will suffice to divide the resource 
fairly---independently of
how the players may value distinct parts of the resource
in a particular case and independently of the strategies chosen
by the players.
In
contrast, in finite \emph{unbounded}
cake-cutting protocols, we cannot predict an
upper bound on how many steps will be required to achieve the same goal.
Aiming to apply cake-cutting procedures to real-world scenarios, it is
important to develop fair \emph{finite bounded} cake-cutting protocols.
In this context, ``fairness'' is often interpreted
as meaning ``envy-freeness.''  A division is \emph{envy-free} if
no player has an incentive to switch his or her portion with the portion
any other player received.
Steinhaus~\cite{ste:j:division-pragmatique}
proved that for any number of players an envy-free division 
of a single divisible good
always exists.
However, the current state of
the art---after six decades of intense
research---is that for arbitrary $n$, and even for $n=4$,
the development of \emph{finite bounded} envy-free
cake-cutting protocols still
appears to be out of reach, and a big challenge for future research.
For $n > 3$ players, hardly any envy-free cake-cutting protocol is known,
and the ones that are known are either finite unbounded or
continuous (see, e.g., \cite{bra-tay:j:protocol,rob-web:j:protocol,bra-tay-zwi:c:moving-knife}).
Though,
from an implementation perspective, finite bounded protocols are
the ones that are most desirable.
Recently, Stromquist~\cite{str:j:finite-protocols} has shown that
for more than two players there is no finite cake-cutting protocol
that provides an envy-free division when all portions are required
to consist of contiguous pieces.  

Our goal in this paper is to look for compromises that can be
made with respect to envy-freeness while keeping the
protocol finite bounded: We
propose an approach to evaluate
finite bounded (yet possibly non-envy-free)
cake-cutting protocols with respect to
their ``degree of guaranteed envy-freeness''  (DGEF).
Informally put, this notion provides a
measure of how good such a protocol can approximate the (possibly for
this particular protocol unreachable) ideal of envy-freeness
in terms of the number
of envy-free-relations
that are guaranteed to exist even in the worst case.
\OMIT{
We stress that our approach of approximating envy-freeness
differs from other lines of research that
also deal with approximating fairness.
For example, Lipton et al.~\cite{lip:c:approximately-fair} propose
to seek for minimum-envy allocations of \emph{indivisible}
goods 
in terms of the value difference of the utility
functions of envied players,
and Edmonds and Pruhs~\cite{edm-pru:c:not-a-piece-of-cake,edm-pru:c:balanced-allocations-of-cake}
approximate fairness in cake-cutting protocols
by allowing merely approximately fair pieces 
(in terms of their value to the players) and by using only approximate
cut queries (in terms of exactness).  Further related approaches will be
mentioned in Section~\ref{sec:def-guaranteed-envy-freeness}, and we
will point out in which ways 
they differ from our approach.
}

This paper is organized as follows.  After defining
some basic notions in
Section~\ref{sec:preliminaries}, we
introduce the notion of
degree of guaranteed envy-freeness, and
specify the DGEF for some well-known
finite bounded proportional cake-cutting protocols
in Section~\ref{sec:guaranteed-envy-freeness}.
In 
Section~\ref{sec:improved-algorithm-guaranteed-envy-freeness}
we present
a new finite bounded proportional cake-cutting protocol with an
enhanced degree of guaranteed envy-freeness, compared with the
proportional protocols mentioned in
Section~\ref{sec:guaranteed-envy-freeness}.  This new cake-cutting
protocol makes use of parallelization in order to include
as many matching valuations (in terms of not raising envy)
as possible. Section~\ref{sec:survey}
briefly describes 
those protocols mentioned in Section~\ref{sec:guaranteed-envy-freeness},
and determines their degree of
guaranteed envy-freeness via a detailed analysis.
In Section~\ref{sec:discussion}, we compare the DGEF approach 
with related work, and show that even small steps toward the 
development of cake-cutting protocols with an enhanced DGEF
are of significance.  Finally, we
conclude in Section~\ref{sec:conclusion} that our approach extends the
scope for the development of new finite bounded cake-cutting protocols
by ``approximating'' envy-freeness instead of insisting on it.

\section{Preliminaries and Notation}
\label{sec:preliminaries}

Cake-cutting is about
dividing a cake into portions that are assigned to
the players such
that each of
them feels, according to his or her individual
valuation of the portions, to have received a fair amount of the
cake.\footnote{As is common, ``cake'' will be used as a metaphor for
the resource or the good to be
divided.}
The cake is assumed to be infinitely divisible, and can be
divided into arbitrary pieces without losing any of its
value. Moreover, we assume the cake to be 
heterogeneous.\footnote{Consider,
for example, a cake with cherry, chocolate, and strawberry toppings.
A player may value, say, the pieces with strawberry
toppings higher than those with cherry toppings.}
This assumption can be made without loss of generality, as any
cake-cutting protocol providing a ``fair'' division of a heterogeneous
resource can be applied in the same way to a homogeneous one
\cite{bra-tay:b:fair-division}.  Given $n$ players, cake $C$ is to
be divided into $n$ portions that
are to be distributed among the players
so as to satisfy each of them.
A portion is not necessarily a single piece of cake but
rather can be a collection of disjoint, possibly noncontiguous pieces
of~$C$.  Furthermore, all players  may have different individual 
valuations of the single pieces of the cake.  For example, one player
may prefer the pieces with the chocolate topping, whereas another player
may prefer the pieces with the cherry topping.

More formally, cake $C$ is represented by the unit interval $[0,1]$
of real numbers. By
performing cuts,
$C$ is divided into $m$ pieces $c_k$, $1 \leq k \leq m$:
Each player~$p_i$, $1 \leq i \leq n$, assigns value $v_i(c_k) =
v_{i}(x_k,y_k)$ to piece $c_k \subseteq C$, where $c_k$ is represented
by the subinterval $[x_k,y_k] \subseteq [0,1]$ and $p_i$'s valuation
function $v_i$ maps subintervals of $[0,1]$ to real numbers in $[0,1]$.
We require each valuation function $v_i$ to satisfy the following
properties:
\begin{enumerate}
\item {\emph{Normalization:}} $v_{i}(0,1)=1$.
\item
  {\emph{Positivity:}}\footnote{%
\label{foo:positivity}%
The literature is a bit ambiguous
regarding this assumption.  Some papers require the
players' values for nonempty pieces of cake to be \emph{nonnegative}
(i.e., $v_i(c_k) \geq~0$) instead of positive.  For example,
Robertson and Webb~\cite{rob-web:b:cake-cutting} and
Woeginger and Sgall~\cite{woe-sga:j:complexity-cake-cutting}
require nonnegative values for nonempty pieces of cake, whereas
positive values for such pieces are required by
Brams and Taylor~\cite{bra-tay:b:fair-division}, Brams, Jones, and 
Klamler~\cite{bra-jon-kla:j:minimal-envy}, and 
Weller~\cite{wel:j:fair-division}.}
For all $c_k \subseteq C$, $c_k \neq \emptyset$, we have $v_i(c_k) >~0$.
\item {\emph{Additivity:}} For all $c_k,c_{\ell} \subseteq C$, $c_k \cap
  c_{\ell} = \emptyset$, we have $v_i(c_k)+v_i(c_{\ell}) = 
v_i(c_k \cup c_{\ell})$.
\item {\emph{Divisibility:}}\footnote{Divisibility implies that for
each $x \in [0,1]$, $v_i(x,x)=0$.  That is, isolated points are
valued~$0$, and open intervals have the
same value as the corresponding closed intervals.
\label{foo:divisibility}}
For all $c_k \subseteq C$ and for each~$\alpha$, $0 \leq
\alpha \leq 1$, there exists some $c_{\ell} \subseteq c_k$ such that
$v_i(c_{\ell}) = \alpha \cdot v_i(c_k)$.
\end{enumerate}
Note that, to simplify notation, we write $v_i(x_k,y_k)$ instead of
$v_i([x_k,y_k])$ for intervals $[x_k,y_k] \subseteq [0,1]$.  Due to
Footnote~\ref{foo:divisibility}, no ambiguity can arise.
For each $[x,y] \subseteq [0,1]$, define $\|[x,y]\| = y-x$.
For any real number~$x$, $\lfloor x \rfloor$ denotes the greatest
integer not exceeding~$x$, and $\lceil x \rceil$ denotes the least
integer not smaller than~$x$.

The assumption that $C$ is heterogeneous formally means that
subintervals of $[0,1]$ having equal size can be valued differently by the
same player.  Moreover, distinct players may value
one and the same piece of the cake differently, i.e., their individual
valuation functions will in general be distinct.
Every player knows only the value of (arbitrary) pieces of $C$
corresponding to his or her own valuation function. Players do not have
any knowledge about the valuation functions of other players.

A \emph{division of $C$} is an assignment of disjoint and nonempty
portions~$C_i \subseteq C$, where $C=\bigcup_{i=1}^{n}
C_i=\bigcup_{k=1}^{m} c_k$, to the players
such that each player $p_i$ receives a portion $C_i \subseteq C$ 
consisting of at least one nonempty piece $c_k \subseteq C$.
The goal of a cake-cutting division is to assign
the portions to the players in as fair a way as possible.
There are different interpretations, though, of what ``fair''
might mean. 
To distinguish between different degrees of fairness, the following
notions have been introduced in the literature (see, e.g.,
Robertson and Webb~\cite{rob-web:b:cake-cutting}):
\begin{definition}
\label{def:proportional-strongfair-envyfree-division}
Let $v_1, v_2, \dots, v_n$ be the valuation functions of the $n$
players.  A division of cake $C = \bigcup_{i=1}^{n} C_i$, where
$C_i$ is the $i$th player's portion, is said to be:
\begin{enumerate}
  \item
  \emph{simple fair} (a.k.a.\
   \emph{proportional}) if and only if for each~$i$, $1 \leq i \leq
   n$, we have $v_i(C_i) \geq \nicefrac{1}{n}$;
  \item
  \emph{strong fair} if and only if
  for each~$i$, $1 \leq i \leq n$, we have $v_i(C_i) >
  \nicefrac{1}{n}$;
  \item
  \emph{envy-free} if and only if for
  each $i$ and~$j$, $1 \leq i, j \leq n$, we have $v_i(C_i) \geq v_i(C_j)$.
\end{enumerate}
\end{definition}

A cake-cutting protocol describes an interactive procedure
for obtaining a division of a given cake, without having
any information on the valuation
functions of the players involved.  Each protocol
is characterized by a set of rules and a set of strategies
(see, e.g., Brams and Taylor~\cite{bra-tay:b:fair-division}).
The rules just determine the course of action, 
such as a request to cut the cake, 
whereas the strategies define how to achieve a certain degree of fairness,
e.g., by advising the players where to cut the cake.
If all players obey the protocol, it is guaranteed that
every player receives a ``fair'' portion of the cake.
Cake-cutting protocols are characterized according to the
degree of fairness of the divisions obtained:\footnote{In 
addition to the fairness criteria given
here, other fairness criteria may be reasonable as well
and have been proposed in the literature (see, e.g., 
Robertson and Webb~\cite{rob-web:b:cake-cutting}).}
\begin{definition}
\label{def:proportional-strongfair-envyfree-protocol}
A cake-cutting protocol is said to be \emph{simple fair} (or
\emph{proportional}), \emph{strong fair}, and \emph{envy-free},
respectively, if every division obtained (i.e., regardless of which
valuation functions the players have) is simple fair (or
proportional), strong fair, and envy-free, respectively, provided that
all players follow the rules and strategies of the protocol.
\end{definition}

Apparently, every division obtained by either a strong fair cake-cutting
protocol or by an envy-free cake-cutting protocol is 
simple
fair as well (i.e., every strong fair or envy-free cake-cutting protocol
can be classified as being simple fair, too).
Moreover,
every simple fair cake-cutting protocol can easily be applied to
the case when there are unequal shares to be assigned, though with
respect to rational ratios only.
Informally speaking, this
can be done by cloning players and their valuation functions so as
to have in total as many players as the smallest common denominator
specifies (see, e.g., \cite{rob-web:b:cake-cutting}).

\section{Degrees of Guaranteed Envy-Freeness for Proportional
Protocols}
\label{sec:guaranteed-envy-freeness}

As mentioned in the introduction, the 
design of envy-free
cake-cutting protocols for any number $n$ of players seems to be
quite a challenge.  For $n \leq 3$~players,
a number of
protocols that always provide
envy-free divisions have been published, both finite (bounded and
unbounded) and continuous ones~\cite{str:j:cut-cake,bra-tay:b:fair-division,rob-web:b:cake-cutting,bar-bra:j:minimal-cuts}.
However,
to the best of our knowledge,
up to date
no finite bounded cake-cutting protocol for $n > 3$~players
is known to always provide an envy-free division.
For practical purposes,
it would be most desirable to have \emph{finite bounded} cake-cutting protocols
that always provide divisions as fair as possible.
In this regard, it is questionable whether
the advantage of always having an envy-free rather than just a proportional
division would be 
big enough to justify the lack of finite boundedness.
It 
may be
worthwhile to be content with a certain lower degree of envy-freeness,
rather than insisting on complete envy-freeness, for the benefit of
having a finite bounded protocol in exchange.

In this paper, we 
propose an approach that
weakens
the concept of
envy-freeness for the purpose of
keeping 
protocols finite bounded.  On the one hand, in
Section~\ref{sec:survey-guaranteed-envy-freeness} we 
are concerned with
known simple fair (i.e., proportional) cake-cutting protocols
that
are finite bounded, and determine their
degree of guaranteed envy-freeness, a notion to be introduced in
Section~\ref{sec:def-guaranteed-envy-freeness} (see
Definition~\ref{def:degree-guaranteed-envy-freeness}).  On the other hand, 
in Section~\ref{sec:improved-algorithm-guaranteed-envy-freeness} we
propose a new finite bounded proportional cake-cutting
protocol that---compared with the known
protocols---has an enhanced degree of guaranteed envy-freeness.

\subsection{Degrees of Guaranteed Envy-Freeness}
\label{sec:def-guaranteed-envy-freeness}

When investigating the degree of envy-freeness of a cake-cutting
protocol for $n$~players, for each player $p_i$, $1 \leq i \leq n$,
the value of his or her portion needs to be compared to the values of
the $n-1$ other portions (according to the measure of 
player~$p_i$).\footnote{We will use ``valuation'' and ``measure''
interchangeably.}
Thus, $n(n-1)$ pairwise relations need to be investigated in order to
determine the degree of envy-freeness of a cake-cutting protocol for $n$
players.  A player $p_i$ envies another player $p_j$, $1 \leq i,j \leq
n$, $i \neq j$, when
$p_i$ prefers player $p_j$'s portion to his
or her own.  If $p_i$ envies~$p_j$,
we call the relation
from $p_i$ to $p_j$ an \emph{envy-relation};
otherwise, we call it an \emph{envy-free-relation}.

\begin{definition}\upshape
\label{def:envy-relation-envy-free-relation}
Consider a division of cake $C=\bigcup_{i=1}^{n} C_i$ for a set $P =
\{p_1,p_2,\dots,p_n\}$ of players, where $v_i$ is $p_i$'s valuation
function and $C_i$ is $p_i$'s portion.
\begin{enumerate}
\item An \emph{envy-relation for this division} (denoted by $\Vdash$)
  is a binary relation on~$P$. Player $p_i$ envies player $p_j$, $1
  \leq i,j \leq n$, $i \neq j$, if and only if $v_i(C_i) <
  v_i(C_j)$. We write $p_i \Vdash p_j$.

\item An \emph{envy-free-relation for this division} (denoted by
  $\nVdash$) is a binary relation on~$P$. Player $p_i$ does not envy
  player $p_j$, $1 \leq i,j \leq n$, $i \neq j$, if and only if
  $v_i(C_i) \geq v_i(C_j)$.  We write $p_i \nVdash p_j$.
\end{enumerate}
\end{definition}

The following properties of envy-relations and envy-free-relations are
worth mentioning.\footnote{Various analogs of envy-relations and
envy-free-relations have also been studied, from an economic perspective,
in the different context of multiagent allocation of indivisible resources.
In particular, Feldman and Weiman~\cite{fel-wei:j:envy-and-wealth}
consider ``non-envy relations'' (which
are similar to our notion of
envy-free-relations) and mention that these are not necessarily transitive.
Chauduri~\cite{cha:j:interpersonal-envy} introduces ``envy-relations''
and mentions that these are irreflexive 
and not necessarily
transitive.  Despite some similarities, their notions differ from ours,
both in their properties and in the way properties holding for their and
our notions are proven.  For example,
Chauduri~\cite{cha:j:interpersonal-envy} notes that mutual envy cannot
occur in a market equilibrium, i.e., in this case his ``envy-relations'' are 
asymmetric, which is in sharp contrast to two-way envy being allowed for our 
notion.}
No player can envy him- or herself, i.e.,
envy-relations are irreflexive: The inequality $v_i(C_i) <
v_i(C_i)$ never holds.
Thus, we trivially have that $v_i(C_i) \geq v_i(C_i)$
always holds.  However, when counting envy-free-relations for a given
division, we will disregard
these trivial envy-free-relations $p_i \nVdash p_i$, $1 \leq i \leq
n$, throughout the paper.

Furthermore, neither envy-relations nor envy-free-relations need
to be transitive.  This is due to the fact that each player values
every piece of the cake according to his or her own valuation
function. The valuation functions of different players
will be distinct in general.  For example,
given three distinct players $p_i$, $p_j$, and $p_k$ with valuations
$v_i(C_i) < v_i(C_j)$ and $v_j(C_j) < v_j(C_k)$, we have that
$p_i \Vdash p_j$ and $p_j \Vdash p_k$.  However, these valuations do not
provide any information about player $p_i$'s valuation of portion
$C_k$, so we cannot conclude that $p_i \Vdash p_k$.
An analogous argument applies to envy-free-relations.

The above observations imply that 
envy-relations and envy-free-relations are either one-way or
two-way, i.e.,
it is possible that:
\begin{enumerate}
\item two players envy each other ($p_i \Vdash p_j$ and $p_j \Vdash
p_i$), which we refer to as ``two-way envy,''
\item neither of two players envies the other ($p_i \nVdash p_j$ and
$p_j \nVdash p_i$), which we refer to as ``two-way envy-freeness,'' and
\item one player envies another player but is not envied by this other
player ($p_i \Vdash p_j$ and $p_j \nVdash p_i$),
which we refer to as both ``one-way envy'' 
(from $p_i$ to~$p_j$) 
and ``one-way envy-freeness'' 
(from $p_j$ to~$p_i$).
\end{enumerate}

Assuming that all players are following the rules and strategies,
some cake-cutting protocols always
guarantee an envy-free division (i.e., they always find an
envy-free division of the cake), whereas others do not. Only
protocols that \emph{guarantee} an envy-free division in \emph{every} case,
even in the worst case (in terms of the players' valuation functions),
are considered to be envy-free.
Note that an envy-free division may
be obtained by coincidence, just because the players have matching
valuation functions that avoid envy, and not because
envy-freeness is enforced by the rules and
strategies of the cake-cutting protocol used.
In the worst case, however, when the players have totally
nonconforming valuation functions, an envy-free division would not
just happen by coincidence, but needs to be enforced by the rules and
strategies of the protocol.
An envy-free-relation is said to be \emph{guaranteed} if it exists even in the
worst case.

We now define the degree of guaranteed
envy-freeness in relation to the problem of cake-cutting.

\begin{definition}\upshape
\label{def:degree-guaranteed-envy-freeness}
For $n \geq 1$ players, the \emph{degree of guaranteed envy-freeness}
(DGEF, for short) of a given 
proportional\footnote{The
DGEF 
should be restricted to proportional protocols only, since otherwise
the DGEF may overstate the actual
level of fairness, e.g., if all the cake is given
to a single player.}
cake-cutting protocol is defined to be the maximum
number of envy-free-relations that exist in every division
obtained by this protocol (provided that all players follow the rules
and strategies of the protocol), i.e., the DGEF (which is expressed as
a function of~$n$) is the number of envy-free-relations that can be
guaranteed even in the worst case.
\end{definition}

By a slight abuse of notation, we will sometimes speak of the number
of guaranteed envy-free-relations (rather than of the guaranteed
number of envy-free-relations).  When we do so, let us remind the
reader that what matters is the \emph{total number} of
envy-free-relations that exist in the worst case, and not the
\emph{identification} of specific envy-free-relations.
Moreover,
for technical reasons (see the proof of Lemma~\ref{lem:divide-conquer}),
we also consider the case that there is only one player (i.e., $n=1$).
Note, however,
that in this case the DGEF of any cake-cutting protocol is trivially zero,
since we disregard the trivial envy-free-relation
$p_1 \nVdash p_1$.

Definition~\ref{def:degree-guaranteed-envy-freeness} 
is based on the idea of weakening the notion
of fairness in terms of envy-freeness in order to obtain
cake-cutting protocols that are fair (though perhaps non-envy-free)
\emph{and} finite bounded,
where the fairness level of a protocol is given by its degree of
guaranteed envy-freeness. The higher the degree of guaranteed
envy-freeness the fairer the protocol.

\subsection{Degrees of Guaranteed Envy-Freeness in Proportional
 Cake-Cutting Protocols}
\label{sec:survey-guaranteed-envy-freeness}

We now give an upper and a lower bound on the degree of guaranteed
envy-freeness for proportional cake-cutting
protocols.  For comparison, note that Feldman and
Kirman~\cite{fel-kir:j:fairness-and-envy} observed that, for
\emph{any} division, the number of envy-relations is always between
zero and $n(n-1)$; zero if everyone is happy with his or her share of
the cake, and $n(n-1)$ if everyone is envious of everyone else.

\begin{proposition}
\label{prop:DGEF-Minimum-Maximum}
  Let $d(n)$ be the degree of guaranteed envy-freeness of
  a proportional cake-cutting protocol for $n \geq 2$
  players. It holds that $n \leq d(n) \leq n(n-1)$.
\end{proposition}

\begin{proofs}
If $n=2$ then we obviously have $d(2) = 2$, and this will be the case
exactly if both players value their share of the cake as being at
least~$\nicefrac{1}{2}$.  Note that this case, $n = 2 = d(n)$, reflects
the fact that every proportional protocol for two players is envy-free.

So, we now assume that $n \geq 3$.
As stated above, we
disregard the trivial envy-free-relation
$p_i \nVdash p_i$ for each~$i$, $1 \leq i \leq n$. Consequently, each player
can have at most $n-1$ envy-free-relations, one to each of the other 
players, which
gives a total of at most $n(n-1)$ guaranteed
envy-free-relations. This proves the upper bound: $d(n) \leq n(n-1)$.

To prove the lower bound, note that, by definition,
in a proportional division every player~$p_i$, 
$1 \leq i \leq n$, regards his or her portion being of value at
least $\nicefrac{1}{n}$, i.e., $v_i(C_i) \geq \nicefrac{1}{n}$.  Thus,
$v_i(C-C_i) \leq \nicefrac{(n-1)}{n}$ for each~$i$.
We will now prove that this
implies that
none of the players can envy each of the $n-1$ other players
at the same time.
We show this for player~$p_1$;
the argument is analogous for the other players.
So, assume that $p_1$ envies some other player, say~$p_2$.  Thus
$v_1(C_2) > v_1(C_1) \geq \nicefrac{1}{n}$.
It follows that $p_1$ values the remaining cake
$(C-C_1)-C_2$ as being less than $\nicefrac{(n-2)}{n}$, i.e.,
$v_1((C-C_1)-C_2) < \nicefrac{(n-2)}{n}$.
Consequently, there is no way
to divide the remaining cake $(C-C_1)-C_2$ into $n-2$ portions
$C_3, C_4, \dots, C_n$
such that
for each $i$, $3 \leq i \leq n$, we have
$v_1(C_i) \geq \nicefrac{1}{n}$.
Hence, there must be at least one player~$p_j$, $3 \leq
j \leq n$, such that $v_1(C_j) < \nicefrac{1}{n} \leq v_1(C_1)$,
so $p_1 \nVdash p_j$.
Considering all $n$ players, this
gives at least $n$ guaranteed envy-free-relations for a proportional
protocol, so $d(n) \geq n$.~\end{proofs}

The degree of fairness of a division obtained by applying a proportional
cake-cutting protocol highly depends on the rules of this protocol.
Specifying and committing to appropriate rules often increases
the degree of guaranteed envy-freeness, whereas the lack of
such rules jeopardizes
it in the sense that the number of guaranteed envy-free-relations may
be limited to the worst-case minimum of $n$ as stated in
Proposition~\ref{prop:DGEF-Minimum-Maximum}.
In this context, ``appropriate rules'' are those that involve
the players' evaluations of other players' portions and of pieces 
that still are to be assigned.  
Concerning a particular piece of cake, involving the evaluations
of as many players as possible in the allocation process
helps to keep the number of envy-relations to be created low,
since this allows to determine early on whether a planned allocation
may later turn out to be disadvantageous---and thus allows to take
adequate countermeasures.
In contrast,
omitting mutual evaluations means to forego
additional knowledge that could turn out to
be most valuable later on.

For example, say player~$p_i$
is going to get assigned piece~$c_j$.  If
the protocol asks all other players to evaluate
piece~$c_j$ according to their measures, all envy-relations to be created
by the assignment of piece~$c_j$ to player~$p_i$ can be identified
before the actual assignment and thus countermeasures (such as trimming
piece~$c_j$) can be undertaken.  However, if the protocol requires
no evaluations on behalf of the other players, such envy-relations
cannot be identified early enough to prevent them from happening.

\begin{lemma}
\label{lem:DGEF-no-evaluations}
Let a proportional cake-cutting protocol for 
$n \geq 2$
players be given.
If the rules of the protocol require none of the players to value any of
the other players' portions, then the degree of guaranteed envy-freeness is~$n$
(i.e., each player is guaranteed only one envy-free-relation).
\end{lemma}

\begin{proofs}
Having a certain number of guaranteed envy-free-relations means to have at
least this number in any case, even in the worst case.
For $n=2$, proportionality implies envy-freeness,
so the worst case is the best case.  For $n \geq 3$ players,
consider the following scenario.
Given a division of cake $C=\bigcup_{i=1}^{n}C_i$,
without any
restrictions other than aiming at a proportional division (i.e.,
the rules of the protocol require none of the players to value
any of the other players' portions), we set the
valuation functions of the players as follows.
For each $i$, $1 \leq i \leq n$,
player $p_i$ values portion $C_i$ to be worth exactly
$\nicefrac{n}{n^2} = \nicefrac{1}{n}$, $p_i$ values exactly
one portion~$C_j$, $i \neq j$, to be worth exactly
$\nicefrac{2}{n^2} < \nicefrac{1}{n}$, and $p_i$ values each of the
$n-2$ remaining portions~$C_k$, $\|\{i,j,k\}\| = 3$, to be worth exactly
$\nicefrac{(n+1)}{n^2} > \nicefrac{1}{n}$.
These valuations make this division 
proportional, as every player values his or her portion to be worth exactly
$\nicefrac{1}{n}$.  Moreover, each player has $n-2$
envy-relations and just one guaranteed envy-free-relation, the latter
of which is due to Proposition~\ref{prop:DGEF-Minimum-Maximum}.
Hence, if the rules of the protocol require none of the players to
value any of the other players' portions, then no more than $n$
envy-free-relations can be guaranteed
by the given proportional cake-cutting protocol
in the worst case.~\end{proofs}

The argument in the proof of Lemma~\ref{lem:DGEF-no-evaluations}
will be used to prove the upper
bounds on the DGEF of the proportional cake-cutting protocols
considered in Theorem~\ref{thm:survey-DGEF} (see also
Lemmas~\ref{lem:last-diminisher}
through~\ref{lem:divide-and-choose}) and
Theorems~\ref{thm:n4-DGEF} and~\ref{thm:n-DGEF}.

An envy-free cake-cutting protocol for $n$ players guarantees that
no player $p_i$ envies
any other player~$p_j$, i.e., $v_i(C_i) \geq v_i(C_j)$ for all $i,j$
with $1 \leq i,j \leq n$,
in each division
$C=\bigcup_{i=1}^{n} C_i$ obtained by this protocol.
This is the case exactly if the protocol meets
the upper bound on the degree of guaranteed envy-freeness
given in Proposition~\ref{prop:DGEF-Minimum-Maximum}.
That is, a cake-cutting protocol for $n \geq 2$ players is
envy-free (or completely fair) exactly if the degree of guaranteed
envy-freeness equals $n(n-1)$.

Our next result shows the DGEF for a number of well-known finite
bounded \emph{proportional} cake-cutting protocols.\footnote{These
protocols may also be known under different names.}
For the sake of self-containment, these protocols will be described in
Section~\ref{sec:survey}.

\begin{theorem}
\label{thm:survey-DGEF}
For $n \geq 3$ players,\footnote{\label{fn:trivial-cases}The trivial
cases $n=1$ (where one player receives all the cake) and $n=2$ 
(where each proportional division is always envy-free) are ignored.
Specifically, an envy-free (and thus proportional) division for $n=2$ 
players can always be obtained by applying the cut-and-choose protocol:
One player cuts the cake into two pieces both of which he or she considers to
be worth exactly one half of the cake, and the other player chooses the
piece that he or she considers to be worth at least half of the cake.}
the proportional cake-cutting protocols listed
in Table~\ref{tab:DGEF-survey} have a degree of guaranteed envy-freeness
as shown in the same table.
\end{theorem}

The proof of Theorem~\ref{thm:survey-DGEF} can be found in
Section~\ref{sec:survey}.  In particular, the proofs of 
Lemmas~\ref{lem:last-diminisher}
through~\ref{lem:divide-and-choose}
provide the details of the analysis yielding the values in the DGEF
column of Table~\ref{tab:DGEF-survey}.

\begin{table}[h!tp]
\centering
  \begin{tabular}{|l||c|c|}
  \hline
  \multicolumn{1}{|c||}{Protocol} & DGEF & Established via \\
  \hline \hline
  Last Diminisher
 \cite{ste:j:steinhaus}
 & $2 + \nicefrac{n(n-1)}{2}$ & Lemma~\ref{lem:last-diminisher}
 \\ \hline
  Lone Chooser
 \cite{fin:j:fair-division}
 & $n$ & Lemma~\ref{lem:lone-chooser} \\ \hline
  Lone Divider
 \cite{kuh:b:games-fair-division} & $2n-2$ & Lemma~\ref{lem:lone-divider}
 \\ \hline
  Cut Your Own Piece (no strategy) \cite{ste:b:math-snapshots}
 & $n$ & Lemma~\ref{lem:cut-your-own-piece} \\
  Cut Your Own Piece (left-right strategy)
 & $2n-2$ & Lemma~\ref{lem:cut-your-own-piece} \\ \hline
  Divide and Conquer \cite{eve-paz:j:cake-cutting}
 &
$n \cdot \left\lfloor \log n \right\rfloor + 
2n - 2^{\left\lfloor \log n \right\rfloor + 1}$
 & Lemma~\ref{lem:divide-conquer} \\
   Minimal-Envy Divide and Conquer
 \cite{bra-jon-kla:j:minimal-envy}
 &
$n \cdot \left\lfloor \log n \right\rfloor + 
2n - 2^{\left\lfloor \log n \right\rfloor + 1}$
 & Lemma~\ref{lem:divide-conquer} \\ \hline
 Recursive Divide and Choose
 \cite{tas:j:proportional-protocol}
 & $n$ & Lemma~\ref{lem:divide-and-choose} \\ 
  \hline
  \end{tabular}
  \caption{DGEF of selected finite bounded cake-cutting
protocols.}
  \label{tab:DGEF-survey}
\end{table}

Apparently, the 
degrees of guaranteed envy-freeness
of the protocols listed in
Table~\ref{tab:DGEF-survey} 
vary
considerably.
Although each of the protocols may provide
an envy-free division in the best case, in the worst case some of them
show just the minimum number of guaranteed envy-free-relations according
to Proposition~\ref{prop:DGEF-Minimum-Maximum}, while others possess
a significantly higher degree of guaranteed envy-freeness.
These differences can be
explained by the fact that these
protocols have been developed with a focus on achieving
proportionality, and not
on maximizing the degree of guaranteed envy-freeness. However, this indicates
a new direction for future research, namely to increase the number of
guaranteed envy-free-relations while ensuring 
finite boundedness.
In the next section, we take a first step in this direction.

\section{Enhancing the Degree of Guaranteed Envy-Freeness}
\label{sec:improved-algorithm-guaranteed-envy-freeness}

In this section, we introduce a finite bounded
cake-cutting protocol that, compared with the protocols in
Table~\ref{tab:DGEF-survey}, improves upon the
degree of guaranteed envy-freeness.
We will prove that this protocol is
proportional and strategy-proof, and that
it can be adapted so as to even
provide a strong fair division.
To present the protocol and its properties in an accessible
way, we first handle the
case of $n=4$ players separately
in Section~\ref{sec:improved-algorithm-n-4}, before
presenting and analyzing it for arbitrary $n \geq 3$ in
Section~\ref{sec:improved-algorithm-arbitrary-n}.  

\subsection{A Proportional Protocol with an Enhanced DGEF for
Four Players}
\label{sec:improved-algorithm-n-4}

Figure~\ref{algo:n4}
gives a finite bounded proportional cake-cutting protocol for four
players.
We give both the rules and the strategies at
once.  Note that the players have to follow the rules and strategies in
order to obtain a proportional share of the cake in any case.

The protocol
in Figure~\ref{algo:n4} always provides a
proportional division (see Theorem~\ref{thm:n4-proportional}) and has
ten guaranteed envy-free-relations (see Theorem~\ref{thm:n4-DGEF}).
To allow comparison for four players, the
best DGEF of any of the proportional protocols
listed in
Table~\ref{tab:DGEF-survey} (see also Section~\ref{sec:survey})---namely,
that of both the Divide and Conquer protocols and the Last
Diminisher protocol---is
eight.
Note that a maximum number of
twelve guaranteed envy-free-relations is possible for four players
(and this would give envy-freeness). 
Moreover, the protocol can be proven
to be strategy-proof (see Theorem~\ref{thm:n4-strategy-proof}),
and to even yield a strong fair division, provided
that exactly one player makes a mark in Step~1 of the protocol
that is closest to~$1$ (see Theorem~\ref{thm:n4-strong-fair}).

\begin{figure}[h!tp]
\centering
\begin{tabular}{|lp{118mm}|}
\hline
{\bf Input:} & Cake~$C$, and four players $p_1, p_2, p_3, p_4$, where $p_i$
has the valuation function~$v_i$.
Note that the value of cake~$C$ is normalized such that 
$v_i(C)=1$, $1 \leq i \leq 4$. \\
{\bf Output:} & Mapping of portions $C_i$ to players $p_i$, where
$C=\bigcup_{i=1}^{4} C_i$. \vspace*{3mm} \\
{\bf Step~1.} & Let each player~$p_i$, $1 \leq i \leq 4$, make a mark
at $m_i \in C$, such that $v_i(m_i,1)=\nicefrac{1}{4}$. \\
{\bf Step~2.} & Find any player $p_j$ such that there is no player $p_k$,
$1 \leq j,k \leq 4$, $j \neq k$, with $\|[m_k,1]\| < \|[m_j,1]\|$. 
(Ties can be broken arbitrarily.) \\
{\bf Step~3.} & Assign portion $C_j=[m_j,1]$ to player $p_j$ and let
$p_j$ drop out. \\
\multicolumn{2}{|l|}{Denote the remaining players by $p_1$, $p_2$,
and $p_3$, without loss of generality, and let $v_i$ be} \\
\multicolumn{2}{|l|}{$p_i$'s valuation function.} \\
{\bf Step~4.} & Let player
$p_1$ cut $[0,m_j]$ into three pieces, say
$c_x$, $c_y$, and $c_z$, of equal
value according to~$v_1$. \\
{\bf Step~5.} &
If $v_2(c_x) > v_2(c_y)$ and
$v_2(c_x) > v_2(c_z)$,
$\{x, y, z\} = \{1, 2, 3\}$, let
player $p_2$ trim piece $c_x$ into piece $c'_x$ and trimmings $R$ such
that
$v_2(c'_x) = v_2(c_y) \geq v_2(c_z)$ or 
$v_2(c'_x) = v_2(c_z) \geq v_2(c_y)$. If there
already exists a two-way tie for the
most valuable piece according to~$v_2$,
do nothing. \\
{\bf Step~6.} & Let player $p_3$ choose one
of the three pieces $c_x$ (respectively, $c'_x$ if $c_x$ has been trimmed),
$c_y$, or $c_z$ that is most valuable according to~$v_3$. \\
{\bf Step~7.} & Let player $p_2$ choose one piece from the two
remaining pieces that is most valuable according to~$v_2$. If
$c'_x$ is
among the remaining
pieces, player $p_2$ has to choose this one. \\
{\bf Step~8.} & Assign the remaining piece to player $p_1$. \\
\multicolumn{2}{|l|}{If trimmings have been made in Step~5, continue
with Step~9, otherwise finished.} \\
{\bf Step~9.} & Either player $p_2$ or player $p_3$ received the
trimmed piece $c'_x$. From these two, let the player not having
received $c'_x$ cut $R$ into three equal pieces according to his or
her measure. \\
{\bf Step~10.} & Let the player having received $c'_x$ choose one
of these three pieces that is most valuable according to his or her
measure. \\
{\bf Step~11.} & Let player $p_1$ choose one out of the two remaining
pieces that is most valuable according to~$v_1$. \\
{\bf Step~12.} & Assign the remaining piece to the player that cut $R$. \\
\hline
\end{tabular}
\caption{A finite bounded proportional cake-cutting protocol
with DGEF of $10$ for
four players.}
\label{algo:n4}
\end{figure}

Steps~4 through~12 of the protocol in Figure~\ref{algo:n4} (which is
the part of the protocol when there are just three players left) is
simply the Selfridge--Conway protocol.\footnote{The Selfridge--Conway
protocol is known to be a finite bounded envy-free
cake-cutting protocol for $n=3$ players (see
Stromquist~\cite{str:j:cut-cake} and also, e.g.,
\cite{bra-tay:j:protocol,bra-tay:b:fair-division,rob-web:b:cake-cutting}).}
We explicitly describe the Selfridge--Conway protocol here for the
sake of self-containment.  Note that the Selfridge--Conway
protocol is also part of the
more involved protocol for an arbitrary number of players, which will
be presented as Figure~\ref{algo:n} in
Section~\ref{sec:improved-algorithm-arbitrary-n}.

\begin{theorem}
\label{thm:n4-proportional}
  The cake-cutting protocol in Figure~\ref{algo:n4} is proportional.
\end{theorem}

\begin{proofs}
 Since all three players entering Step~4 consider the portion
 of the fourth player (who dropped out in Step~3)
 as being worth no more than~$\nicefrac{1}{4}$,
 the Selfridge--Conway protocol is applied to a part of
 the cake that all three involved players consider as being worth at
 least $\nicefrac{3}{4}$. Thus, since the Selfridge--Conway protocol is
 an envy-free (hence, in particular, a proportional)
 protocol, it by definition guarantees each of the three
 players entering Step~4 a portion of
 value at least $\nicefrac{1}{4}$
 according to their measures. Moreover, the player
 who dropped out in Step~3
 considers his or her portion to be worth $\nicefrac{1}{4}$ due
 to Step~1. Therefore, the cake-cutting protocol
 in Figure~\ref{algo:n4} always provides a proportional division
 for $n=4$ players.~\end{proofs}

\begin{theorem}
\label{thm:n4-DGEF}
  The cake-cutting protocol in Figure~\ref{algo:n4} has
  ten guaranteed
  envy-free-relations.
\end{theorem}

\begin{proofs}
The DGEF of the protocol in Figure~\ref{algo:n4}
can be justified analogously to the
arguments in the proof of Theorem~\ref{thm:n4-proportional}.
Because the
Selfridge--Conway protocol always provides an envy-free division, there is no
envy between the three players entering Step~4, which results in
six
guaranteed envy-free-relations. In addition, the same three players
will not envy the player (call him or her~$p_j$)
who dropped out in Step~3 with portion~$C_j$,
since none of
them valued portion $C_j$ as being worth more than~$\nicefrac{1}{4}$
and each of them is guaranteed a share of at least
$(\nicefrac{1}{3})(\nicefrac{3}{4}) = \nicefrac{1}{4}$, 
which gives
three more guaranteed envy-free-relations.
Simply put, none of the three players entering Step~4 will envy 
any of the three other players, summing up to nine guaranteed 
envy-free-relations.  By the argument in the 
proof of Proposition~\ref{prop:DGEF-Minimum-Maximum},
no more envy-free-relations can be 
guaranteed on behalf of these three players.
The last guaranteed
envy-free-relation
is due to player~$p_j$:
Since the protocol always provides a proportional division, $p_j$
cannot envy each of the three remaining players  
(again by the proof of
Proposition~\ref{prop:DGEF-Minimum-Maximum}).
However, it cannot be guaranteed that 
$p_j$ does not envy any of the other two remaining players either,
since $p_j$ does not evaluate their portions.  This follows from
the proof of Lemma~\ref{lem:DGEF-no-evaluations}.\footnote{More
specifically, consider the following scenario.  Suppose the fourth player 
dropped out in Step~3 (so $j=4$), and $p_4$ values
his or her portion $C_4$ to be worth exactly~$\nicefrac{1}{4}$
and the remaining
cake $[0, m_4]$ to be worth exactly~$\nicefrac{3}{4}$. If we set the
valuation functions
such that $v_1(C_1)=v_2(C_2)=v_3(C_3)=\nicefrac{1}{4}$,
$v_4(C_1)=\nicefrac{1}{8}$, and $v_4(C_2)=v_4(C_3)=\nicefrac{5}{16}$
(thus, $v_4(C_1)+v_4(C_2)+v_4(C_3)=\nicefrac{3}{4}$),
dividing cake $C$ into $C_1$, $C_2$, $C_3$, and $C_4$
results in a
proportional division
such that player $p_4$ envies players $p_2$ and~$p_3$.}
Thus, the protocol
shown in Figure~\ref{algo:n4} has a DGEF of exactly
ten in total.~\end{proofs}

Thinking of manipulation aspects, there may be players who
try to gain most of the cake for themselves, or who intentionally
try to make other players envious.  To prevent this from happening,
cake-cutting protocols should be strategy-proof.

\begin{definition}
\label{def:proportional-strategy-proof}
  A proportional cake-cutting protocol is said to be \emph{strategy-proof}
  if a cheating player is no longer guaranteed a proportional share,
  whereas all other players are still guaranteed to receive their
  proportional share.
\end{definition}

In a strategy-proof proportional protocol, a cheater (i.e., a player
who doesn't play truthfully) cannot harm any of the other players with
respect to proportionality, and may even jeopardize receiving a
proportional share of the cake for him- or herself.

It is worth noting that
the definition of strategy-proofness is slightly stronger when
restricted to \emph{envy-free} cake-cutting protocols.
An envy-free cake-cutting protocol is said to be strategy-proof
if a cheating player is no longer guaranteed to not envy any other player,
whereas all other players are.
That is, a strategy-proof envy-free cake-cutting protocol is resistant
to manipulation in the sense that for a player to be \emph{guaranteed}
to not envy any other player, he or she is required to play truthfully.

\begin{theorem}
\label{thm:n4-strategy-proof}
  The proportional cake-cutting protocol in Figure~\ref{algo:n4} is
  strategy-proof in the sense of
  Definition~\ref{def:proportional-strategy-proof}.
\end{theorem}

\begin{proofs}
When analyzing the strategy-proofness of the protocol
in Figure~\ref{algo:n4}, only
decisions made in Steps~1 through~3 need to be considered, since the
Selfridge--Conway protocol has already been proven to be
strategy-proof (see, e.g., \cite{bra-tay:b:fair-division}).
So, in each of the
following three cases,
we thus will consider only Steps~1 through~3
of the protocol.  Moreover, it is assumed that there is exactly one
cheater (i.e., one player not playing truthfully) that is trying to get
more than a proportional share,
call him or her~$p_c$, $1 \leq c \leq 4$.

\begin{fall}
\item \label{n4-strategy-proof-case1}
  If $p_c$ is the player to drop out in Step~3 
  with portion $C_c=[m_c,1]$, $v_c(m_c,1) > \nicefrac{1}{4}$,
  then the cheater would receive
  more than a proportional share according to his or her measure.
  However, all other players are still
  guaranteed a proportional share, since all of them consider
  portion $C_c$ as being worth at most~$\nicefrac{1}{4}$
  according to their measures, so they all consider the remaining
  part of the cake to be worth at least~$\nicefrac{3}{4}$.
\item \label{n4-strategy-proof-case2}
  If the cheater, $p_c$, would
  have dropped out in Step~3 with portion $C_c=[m_c,1]$ when telling the truth
  (i.e., $v_c(m_c,1)=\nicefrac{1}{4}$ and $v_i(m_c,1) \leq \nicefrac{1}{4}$
  with $1 \leq i \leq 4$ and $i \neq c$)
  but now is not (since $p_c$ makes a mark at $m_c^\prime$ with
  $m_c^\prime < m_c$),
  then the cheater
  may end up with even less 
  than~$\nicefrac{1}{4}$. Let us see why this is true.  In this case,
  another player, $p_j$, drops out in Step~3 with
  portion $C_j=[m_j,1]$, where $m_c^\prime \leq m_j \leq m_c$, 
  which determines the subpart of the
  cake for Step~4 and the following steps to be $[0,m_j]$. According to the
  measure of the cheater, subpart $[0,m_j]$ is worth at
  most~$\nicefrac{3}{4}$, since he or she values $[m_j,1]$ as being
  worth at least
  $\nicefrac{1}{4}$, as assumed beforehand. Applying the Selfridge--Conway
  protocol to subpart $[0,m_j]$ guarantees each of the involved
  players at least a proportional share, which results for the cheater in a
  portion that may be worth even
  less than $(\nicefrac{1}{3})(\nicefrac{3}{4}) = \nicefrac{1}{4}$ (namely,
  if $m_j < m_c$).  (Note that this is the point where the cheater loses
  his or her guarantee for a proportional share via cheating.)

  Again, all other players are
  still guaranteed a proportional share. The player receiving portion
  $C_j=[m_j,1]$ values this portion as being~$\nicefrac{1}{4}$.
  The two remaining players
  both value subpart $[0,m_j]$ to be worth
  at least~$\nicefrac{3}{4}$, and thus
  receive a portion that is worth at least
  $(\nicefrac{1}{3})(\nicefrac{3}{4})$ according to their measures.
\item \label{n4-strategy-proof-case3}
  If the cheater does
  not drop out in Step~3 by cheating, nor would have dropped out in Step~3
  if he or she would have been truthfully,
  then this would not influence
  the division at all. The player dropping out in Step~3 with
  portion $C_j=[m_j,1]$
  values this portion as being~$\nicefrac{1}{4}$,
  and the remaining players receive a
  proportional share of subpart $[0,m_j]$, which all of them value at
  least $\nicefrac{3}{4}$, even the cheater.
 \end{fall}
This concludes the proof of
Theorem~\ref{thm:n4-strategy-proof}.~\end{proofs}

Moreover, just a little change in the procedure can make the protocol
presented above to provide not just a simple fair but a strong fair
division for four players, 
which means that every player considers the portion received to
be worth strictly more than
one quarter of $C$.

\begin{theorem}
\label{thm:n4-strong-fair}
The cake-cutting protocol in Figure~\ref{algo:n4} can be modified so
as to yield a strong fair division, provided that
exactly one player makes a mark in Step~1 that is closest to~$1$
(with respect to the interval $[0,1]$).
\end{theorem}

\begin{figure}[h!tp]
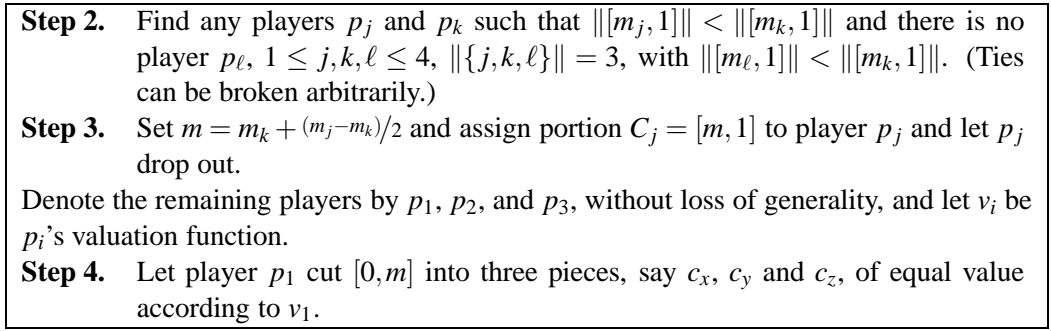

\centering
\begin{tabular}{|lp{118mm}|}
\hline 
{\bf Step~2.} & Find any players $p_j$ and $p_k$ such that
$\|[m_j,1]\| < \|[m_k,1]\|$ and there is no player $p_{\ell}$, $1 \leq
j,k,\ell \leq 4$, $\|\{j,k,\ell\}\|=3$, with $\|[m_{\ell},1]\| < \|[m_k,1]\|$.
(Ties can be broken arbitrarily.) \\
{\bf Step~3.} & Set $m=m_k+\nicefrac{(m_j-m_k)}{2}$ and assign portion
$C_j=[m,1]$ to player $p_j$ and let $p_j$ drop out. \\
\multicolumn{2}{|l|}{Denote the remaining players by $p_1$, $p_2$,
and~$p_3$, without loss of generality, and let $v_i$ be} \\
\multicolumn{2}{|l|}{$p_i$'s valuation function.} \\
{\bf Step~4.} & Let player $p_1$ cut $[0,m]$ into three pieces, say $c_x$,
$c_y$ and $c_z$, of equal
value according to~$v_1$. \\
\hline
\end{tabular}
\caption{Modified steps in the protocol of Figure~\ref{algo:n4} 
to achieve a strong fair division for
four players.}
\label{algo:n4:strongfair}
\end{figure}

\begin{proofs}
Figure~\ref{algo:n4:strongfair} shows only the modified steps of the
protocol in Figure~\ref{algo:n4} that are required to achieve a
strong fair division.

Let $p_j$ be the unique player whose mark in Step~1 is closest to~$1$.
According to Step~2 in Figure~\ref{algo:n4:strongfair}, let $p_k$ be
any player such that $\|[m_j,1]\| < \|[m_k,1]\|$ and there is no
player $p_{\ell}$, $1 \leq j,k,\ell \leq 4$, $\|\{j,k,\ell\}\|=3$,
with $\|[m_{\ell},1]\| < \|[m_k,1]\|$, where ties can be broken
arbitrarily.  Step~3 in Figure~\ref{algo:n4:strongfair} assures that
player $p_j$ is dropping out with a portion that is worth strictly
more than $\nicefrac{1}{4}$ according to his or her
measure,\footnote{Recall that we assumed the axiom of positivity,
which requires nonempty pieces of cake to have a nonzero value for
each player, see also Footnote~\ref{foo:positivity}.} since $p_j$
receives a portion that is bigger and thus is worth more than the one he
or she has marked as being worth exactly~$\nicefrac{1}{4}$. The three
remaining players continue by applying a proportional protocol to a
part of the cake that all of them consider to be worth strictly more
than~$\nicefrac{3}{4}$. Thus,
each of the players receives a portion that is worth strictly more
than $\nicefrac{1}{4}$ according to his or her measure, which results
in a strong fair division.~\end{proofs}

\subsection{A Proportional Protocol with an Enhanced DGEF for
any Number of Players}
\label{sec:improved-algorithm-arbitrary-n}

Figure~\ref{algo:n} shows a finite bounded proportional cake-cutting protocol
with an enhanced DGEF for $n$ players, where 
$n \geq 3$
is arbitrary.
Again, we give both the rules and the strategies at once, and players
have to follow the rules and strategies in order to obtain 
a proportional share of the cake.
Unless specified otherwise, ties in this protocol can be broken
arbitrarily.
With respect to the DGEF results of previously known
finite bounded
proportional cake-cutting protocols given in Table~\ref{tab:DGEF-survey}
(see also Section~\ref{sec:survey}),
the Last Diminisher protocol\footnote{This protocol
has been developed by Banach and Knaster
and was first presented 
in
Steinhaus~\cite{ste:j:steinhaus}.}
shows the best results for $n \geq 6$, whereas
the best results for $n < 6$ are achieved by
the Last Diminisher protocol as well as both the 
Divide and Conquer protocols
\cite{eve-paz:j:cake-cutting,bra-jon-kla:j:minimal-envy}.
The protocol presented in Figure~\ref{algo:n} improves upon these
protocols in terms of the degree of 
guaranteed envy-freeness for all $n \geq 3$ and, in particular,
improves upon the DGEF of
the Last Diminisher protocol
by $\left\lceil \nicefrac{n}{2} \right\rceil - 1$ additional
guaranteed envy-free-relations.\footnote{Recall that we ignore the
trivial cases $n=1$ and $n=2$, see
Footnote~\ref{fn:trivial-cases}.  Moreover, in the special case of~$n=4$
the DGEF of the protocol in Figure~\ref{algo:n}
is~$\left( \nicefrac{n^2}{2} \right) + 2 = 10$
and thus improves the DGEF of
the Last Diminisher protocol by even~$\nicefrac{n}{2} = 2$
guaranteed envy-free-relations, see Theorem~\ref{thm:n4-DGEF}.}

Both the protocol in Figure~\ref{algo:n} and the Last 
Diminisher protocol are, more or less, based on the 
same idea of determining a piece of minimal size that is valued exactly 
$\nicefrac{1}{n}$ by one of the players (who is still in the game),
which guarantees that all other players (who are
still in the game) will not envy this
player for receiving this particular piece.
However, the protocol in Figure~\ref{algo:n}
works in a more parallel way, which
makes its enhanced DGEF of 
$\left\lceil \nicefrac{n^2}{2} \right\rceil + 1$ possible
(see Theorem~\ref{thm:n-DGEF}), and it forbears from using trimmings.
To ensure that working in a parallel manner indeed pays off in terms of 
increasing the degree of guaranteed envy-freeness, the ``inner loop''
(Steps~4.1 through~4.3) of the protocol is decisive.

In addition, the protocol in Figure~\ref{algo:n} always provides a proportional 
division (see Theorem~\ref{thm:n-proportional}) in a finite bounded 
number of steps (see Theorem~\ref{thm:n-finite-bounded}), 
it 
can be proven to be strategy-proof
(in the sense of Definition~\ref{def:proportional-strategy-proof}, see
Theorem~\ref{thm:n-strategy-proof}), and analogously to the modification
described in Section~\ref{sec:improved-algorithm-n-4}, this protocol
can
be adjusted to provide a strong fair division for suitable valuation
functions of the players
(see Theorem~\ref{thm:n-strong-fair}).
All of the above properties are shared also by the Last Diminisher protocol 
(see, e.g., \cite{rob-web:b:cake-cutting}).

\begin{remark}
\label{rem:remarksblock}
Some remarks on the protocol in Figure~\ref{algo:n} are in order:
\begin{enumerate}
\item \label{rem:remarksblock-1} 
From a very high-level perspective the procedure is as follows:
The protocol runs over several rounds in each of which
it is to find a player~$p_j$ who takes a portion from the left side
of the cake, and to find a player~$p_k$ who takes a disjoint portion
from the right side of the cake, such that none of the players still
in the game envy~$p_j$ or~$p_k$ (at this, appropriate ``inner-loop handling''
might be necessary, see Figure~\ref{algo:n} for details).
Thereafter, $p_j$ and $p_k$ are to drop out with their portions,
and a new round is started with the remaining cake (which is
being renormalized, see Remarks~\ref{rem:remarksblock-3}
and~\ref{rem:remarksblock-5} below) and
the remaining players.  Finally, the Selfridge--Conway protocol is
applied to the last three players in the game.

\item \label{rem:remarksblock-2} Note that this protocol is applied only
if there are more than two players in total.  If there is
just one player then he or she receives all the cake, and if there are only
two players then the simple cut-and-choose protocol is applied (see
Footnote~\ref{fn:trivial-cases}).

\item \label{rem:remarksblock-3} Regarding $n \geq 5$ players,
if at any stage of our protocol the same player marks both the
leftmost smallest piece and the rightmost smallest piece, the cake may
be split up into two pieces and later on merged again.  To simplify
matters, in such a case the interval boundaries are adapted as well,
which is expressed in Step~8 of Figure~\ref{algo:n}.
Simply put, the two parts of the cake are set next to each other again
to ensure a seamless transition.  This can be done without any loss
in value due to additivity of the players' valuation functions.

\item \label{rem:remarksblock-4} Note that if the inner loop
(Steps~4.1 through~4.3) has
not been executed in an outer-loop iteration (Steps~1 through~8), we
have $\varrho = \varrho^{\prime}$.  This is the special case of zero
iterations of the inner loop.  Consequently, if $\varrho =
\varrho^{\prime}$ then the portion $C_k=[\varrho_k,\varrho]$ assigned
to player $p_k$ in Step~6 is the same as
$C_k=[\varrho_k,\varrho^{\prime}]$ in the general case, and the values
for $\varrho:=\varrho_k$ and $C^{\prime}:=[\lambda_j, \varrho_k]$ that
are set in Step~8 in this case are special cases of
$\varrho:=\varrho-\varrho^{\prime}+\varrho_k$ and
$C^{\prime}:=[\lambda_j, \varrho_k] \cup [\varrho^{\prime}, \varrho]$
(since if $\varrho = \varrho^{\prime}$ then $[\varrho^{\prime},
\varrho]$ degenerates to a single point, which is valued zero by the
axiom of divisibility, see Footnote~\ref{foo:divisibility}).  However,
to make the protocol in Figure~\ref{algo:n} easier to comprehend, we
have stated these special cases explicitly in addition to the general
case.

\item \label{rem:remarksblock-5} In Steps~1 and~9.1, the value of
subcake~$C^{\prime} \subseteq C$ is renormalized such that
$v_i(C^{\prime})=1$ for each player $p_i$, $1 \leq i \leq s$,
for the sake of convenience.  In more detail, each player~$p_i$ 
values~$C^{\prime}$ at least $\nicefrac{s}{n}$ of $C$, i.e.,
$v_i(C^{\prime}) \geq (\nicefrac{s}{n}) \cdot v_i(C)$. 
The latter holds true since each of the $s$ players still in the game
values the union of the $n-s$ portions already assigned to be worth
at most $\nicefrac{(n-s)}{n}$ of $C$.  Thus, by receiving 
a proportional share (valued $\nicefrac{1}{s}$) of $C^{\prime}$ 
each player~$p_i$ is guaranteed at least a proportional share (valued 
$\nicefrac{1}{n}$) of $C$.

\item \label{rem:remarksblock-6} Note that Steps~9.1 through~9.3
of the protocol in
Figure~\ref{algo:n} correspond to Steps~1 through~3 of the protocol in
Figure~\ref{algo:n4}, and that Steps~9.1 through~9.4 are performed
exactly if the initial number $n$ of players is even.
\end{enumerate}
\end{remark}

\begin{figure}[h!tp]
  \centering
  \begin{tabular}{|lp{118mm}|}
    \hline
    {\bf Input:} & Cake~$C$, and $n$ players $p_1, p_2, \dots, p_n$,
where $p_i$ has valuation function~$v_i$. \\
    {\bf Output:} & Mapping of portions $C_i$ to players~$p_i$, where 
$C=\bigcup_{i=1}^{n} C_i$. \\
    {\bf Initialization:} & Set $\lambda:=0$, $\varrho:=1$,
$\varrho^{\prime}:=\varrho$, $s:=n$, and $C^{\prime}:=[0,1]=C$.
\\
    \multicolumn{2}{|p{145mm}|}{While there are more than four players
(i.e., $s>4$), perform the outer loop (Steps~1 through~8).} \\
    {\bf Step~1.} & Let players $p_i$, $1 \leq i \leq s$,
each make two marks at
$\lambda_i$ and $\varrho_i$ with $\lambda_i,\varrho_i \in C^{\prime}$ such that
$v_i(\lambda,\lambda_i)=\nicefrac{1}{s}$
and $v_i(\varrho_i,\varrho)=\nicefrac{1}{s}$; note that $v_i(C^{\prime})=1$
(see Remark~\ref{rem:remarksblock}.\ref{rem:remarksblock-5}). \\
    {\bf Step~2.} & Find any player $p_j$ such that there is no player $p_z$,
$1 \leq j,z \leq s$, $j \neq z$,
with $\|[\lambda,\lambda_z]\| < \|[\lambda,\lambda_j]\|$.
\\
    {\bf Step~3.} & Find any player $p_k$ such that there is no player
$p_z$,
$1 \leq k,z \leq s$, $k \neq z$,
with $\|[\varrho_z,\varrho]\| < \|[\varrho_k,\varrho]\|$.
If more than one player fulfills this condition for~$p_k$, and
$p_j$ is one of them, choose $p_k$ other than $p_j$.\\
    \multicolumn{2}{|p{145mm}|}{If $j \neq k$,
go directly to Step~5,
else repeat the inner loop (Steps~4.1 through 4.3) until $p_j$ and $p_k$
are found such that $j \neq k$, where $p_j$ marks the leftmost smallest piece
and $p_k$ marks the rightmost smallest piece.} \\
    {\bf Step~4.1.} & Set $\varrho^{\prime} := \varrho_k$. \\
    {\bf Step~4.2.} & Let players $p_i$, $1 \leq i \leq s$, each make a mark at
$\varrho_i$ with $\varrho_i \in C^{\prime}$ such that
$v_i(\varrho_i,\varrho^{\prime})=\nicefrac{1}{s}$. \\
    {\bf Step~4.3.} & Find player $p_k$ such that there is no player
 $p_z$, $1 \leq k,z \leq s$, $k \neq z$,
with $\|[\varrho_z,\varrho^{\prime}]\| < \|[\varrho_k,\varrho^{\prime}]\|$.
If more than one player fulfills this condition for~$p_k$, and
$p_j$ is one of them, choose $p_k$ other than $p_j$. \\
    {\bf Step~5.} & Assign portion $C_j=[\lambda,\lambda_j]$ to player $p_j$. \\
    {\bf Step~6.} & If $\varrho = \varrho^{\prime}$, assign
portion $C_k=[\varrho_k,\varrho]$ to player $p_k$, else assign portion
$C_k=[\varrho_k,\varrho^{\prime}]$ to player $p_k$ (see
Remark~\ref{rem:remarksblock}.\ref{rem:remarksblock-4}). \\
    {\bf Step~7.} & Let players $p_j$ and $p_k$ drop out. \\
    {\bf Step~8.} & Adapt the interval boundaries of the remaining cake
$C^{\prime}$ for the round / step to follow: If $\varrho = \varrho^{\prime}$, set
$C^{\prime}:=[\lambda_j, \varrho_k]$ and $\varrho:=\varrho_k$, else set
$C^{\prime}:=[\lambda_j, \varrho_k] \cup [\varrho^{\prime}, \varrho]$ and
$\varrho:=\varrho-\varrho^{\prime}+\varrho_k$ 
(see Remarks~\ref{rem:remarksblock}.\ref{rem:remarksblock-3}
and~\ref{rem:remarksblock}.\ref{rem:remarksblock-4}).
Set $\lambda:=\lambda_j$, $\varrho^{\prime} := \varrho$, and $s:=s-2$. \\
   \multicolumn{2}{|p{145mm}|}{Perform Steps~9.1 through 9.4 if and only if
there are four players (i.e., $s=4$).
If there are three players (i.e., $s=3$), go directly to Step~10.} \\
    {\bf Step~9.1.} & Let each $p_i$, $1 \leq i \leq s=4$,  
make a mark at $\varrho_i \in C^{\prime}$ such that
$v_i(\varrho_i,\varrho)=\nicefrac{1}{s}=\nicefrac{1}{4}$;
note that $v_i(C^{\prime})=1$ (see
Remark~\ref{rem:remarksblock}.\ref{rem:remarksblock-5}).  \\
    {\bf Step~9.2.} & Find any player $p_j$ such that there is no player $p_k$,
$1 \leq j,k \leq s$, $j \neq k$ with
$\|[\varrho_k,\varrho]\| < \|[\varrho_j,\varrho]\|$. \\
    {\bf Step~9.3.} & Assign portion $C_j=[\varrho_j,\varrho]$ to player $p_j$.
Let player $p_j$ drop out. \\
    {\bf Step~9.4.} & Set $\varrho:=\varrho_j$, 
$C^{\prime}:=[\lambda, \varrho]$, and $s:=s-1$. \\
    {\bf Step~10.} & Divide the remaining cake $C^{\prime}$ among the 
$s=3$ remaining players via the Selfridge--Conway protocol
 (as described in Steps~4 through 12 in Figure~\ref{algo:n4}).
\\
    \hline
  \end{tabular}
  \caption{A proportional protocol with an enhanced DGEF
of~$\left\lceil \nicefrac{n^2}{2} \right\rceil + 1$ for $n \geq 3$ players.}
  \label{algo:n}
\end{figure}

\begin{theorem}
\label{thm:n-proportional}
  The cake-cutting protocol in Figure~\ref{algo:n} is proportional.
\end{theorem}

\begin{proofs}
In the case of $n$ being even, 
all four players entering Step~9.1 consider $C^{\prime}$ at this stage 
to be worth at least~$\nicefrac{s}{n} = \nicefrac{4}{n}$ of $C$, 
since each of them values the union
of the $n-4$ portions already assigned to be worth
at most $\nicefrac{(n-4)}{n}$ of $C$.  
The same argument can be applied to the
three players that enter Step~10 in the case of $n$ being odd.
Thus,
from Step~9.1 on, which is the part when there are no more than
four players left,
the protocol provides a proportional division according to
Theorem~\ref{thm:n4-proportional}.  

When there are more than four players in total,
either the first $n-4$ players (if $n$ is even),
or the first $n-3$ players (if $n$ is odd), 
receive a portion they each value to be worth
at least $\nicefrac{1}{n}$ according to Steps~1 and~4.2,
since each of these players receives a portion he or she once
specified to be worth exactly~$\nicefrac{1}{s}$ 
of $C^{\prime}$, while valuing $C^{\prime}$
at least~$\nicefrac{s}{n}$ of $C$ 
(see Remark~\ref{rem:remarksblock}.\ref{rem:remarksblock-5}).

Thus, the cake-cutting protocol in
Figure~\ref{algo:n} always provides a proportional division for any number $n
\geq 3$ of players.~\end{proofs}

\begin{theorem}
\label{thm:n-finite-bounded}
  The cake-cutting protocol in Figure~\ref{algo:n} is finite bounded.
\end{theorem}

\begin{proofs}
  The protocol in Figure~\ref{algo:n}
  has only a finite number of steps---with two loops though.
  The outer loop (which repeats Steps~1 through~8 
  as long as there are more than 
  either four players if $n$ is even, or three players if $n$ is odd)
  is iterated $\nicefrac{(n-4)}{2}$ times if $n$ is even, and is
  iterated $\nicefrac{(n-3)}{2}$ times if $n$ is odd.
  The inner loop (which repeats Steps~4.1 through~4.3 until two distinct
  players are found to which the two outermost pieces are assigned, one
  player receiving that from the present left and the other player
  receiving that from the present right boundary) is iterated at most $s-2$
  times per outer-loop iteration with $s$ players, summing up to at most 
  $\sum_{i=1}^{\left\lceil \nicefrac{(n-4)}{2} \right\rceil}{(n-2i)}$
  iterations of the inner loop in total.
  Let us see why this is true.  For each outer-loop iteration
  with $s$ players, 
  the inner loop 
  is iterated as long as
  the player that marked the leftmost smallest piece
  also marks the rightmost
  smallest piece (with respect to the
  current right boundary $\varrho^{\prime}$)
  and there is no tie with another player for the rightmost smallest piece. 
  With respect to every single player~$p_i$, $1 \leq i \leq s$, 
  when setting $v_i(C^{\prime})=1$, 
  the division of $C^{\prime}$
  into pieces valued $\nicefrac{1}{s}$ each results in $s$
  disjoint pieces.  Two of these $s$ pieces have been identified in Steps~2
  and~3 already and thus only $s-2$ more pieces can be identified.
  At this, let $p_j$ be the player that is to be assigned
  the leftmost smallest piece $[\lambda,\lambda_j]$ in Step~5
  and the only player that marked the rightmost 
  smallest piece $[\varrho_j, \varrho]$. 
  Then there must be a player~$p_k$,
  $k \neq j$, with $v_k(\lambda, \lambda_j) \leq \nicefrac{1}{s}$ and
  $v_k(\varrho_j, \varrho) < \nicefrac{1}{s}$, and there is
  at least one piece $[\varrho_k, \varrho^{\prime}]$
  among the $s-2$ remaining pieces for which it
  is true that $v_k(\varrho_k, \varrho^{\prime})=\nicefrac{1}{s}$ and  
  $v_j(\varrho_k, \varrho^{\prime}) \leq \nicefrac{1}{s}$,
  i.e., for some $\varrho^{\prime}$ it holds that
  $\|[\varrho_k,\varrho^{\prime}]\| \leq 
  \|[\varrho_j,\varrho^{\prime}]\|$.
  Hence, within any
  outer-loop iteration, the inner loop stops after at most $s-2$ iterations.
  Counting in single steps, Steps~1, 2, 3, 5, 6, 7, and~8 each are repeated
  $\left\lceil \nicefrac{(n-4)}{2} \right\rceil$ times,
  Steps~4.1, 4.2 and~4.3 each are repeated
  at most $\sum_{i=1}^{\left\lceil \nicefrac{(n-4)}{2} \right\rceil}{(n-2i)}$
  times, Steps~9.1
  through~9.4 each are repeated at most once, and Step~10 involves at most nine
  more steps (according to Figure~\ref{algo:n4}).
  Thus, in terms of single steps as presented in
  Figure~\ref{algo:n}, the protocol is bounded by
  $(7 \cdot \left\lceil \nicefrac{(n-4)}{2} \right\rceil) + 
   (3 \cdot \sum_{i=1}^{\left\lceil \nicefrac{(n-4)}{2}
   \right\rceil}{(n-2i)}) + 4 + 9$ steps.
  Thus, the protocol in Figure~\ref{algo:n} carries out only
  finitely many operations and is finite bounded.~\end{proofs}

\begin{theorem}
\label{thm:n-DGEF}
  For $n \geq 5$ players, 
  the cake-cutting protocol
  in Figure~\ref{algo:n} has $\left\lceil \nicefrac{n^2}{2}
  \right\rceil + 1$ guaranteed envy-free-relations.\footnote{Note that 
    the same formula holds if~$n=3$, but for the special case of~$n=4$ even
    one more envy-free-relation can be guaranteed 
    (see~Theorem~\ref{thm:n4-DGEF}).}
\end{theorem}

\begin{proofs}
The DGEF of the protocol in Figure~\ref{algo:n}
increases every time a portion is assigned to a player.
Considering Steps~1 through~8 of any outer-loop iteration
with $s$ players, player~$p_j$ (the player receiving the leftmost smallest
piece in Step~5 and dropping out with this portion in Step~7)
will not be envied by any of the $s-1$ other players.
Regarding player $p_k$ (the player receiving the rightmost smallest piece
in Step~6 and dropping out with this portion in Step~7),
two cases need to be considered---the best case
and the worst case.  In the best case, 
player $p_k$ immediately is not the same one as player~$p_j$.
In the worst case, players $p_j$ and $p_k$ are one and the same player
(i.e., $j=k$) in Step~3, and
Steps~4.1 through 4.3 need to be executed.
In this case,
as already mentioned in the proof of Theorem~\ref{thm:n-finite-bounded},
there must be at least one piece $[\varrho_k, \varrho^{\prime}]$ with 
$v_k(\varrho_k, \varrho^{\prime}) = \nicefrac{1}{s}$ and  
$v_j(\varrho_k, \varrho^{\prime}) \leq \nicefrac{1}{s}$,
i.e., for some $\varrho^{\prime}$ it holds that
$\|[\varrho_k,\varrho^{\prime}]\| \leq 
\|[\varrho_j,\varrho^{\prime}]\|$.
Consequently, in both the best and the
worst case the player receiving the rightmost
smallest piece (with respect to 
the current right boundary,
either $\varrho$ or some $\varrho^{\prime}$)
will not be envied by any of the $s-1$ other players, not even by
player~$p_j$.  However, none of the
players $p_j$ and $p_k$ can be guaranteed 
to be not envied by more than $s-1$ players
according to the proof of Lemma~\ref{lem:DGEF-no-evaluations}.

Since the outer loop is repeated
$\left\lceil \nicefrac{(n-4)}{2} \right\rceil$ times, the number of
guaranteed envy-free-relations among the players
sums up to $\left(\nicefrac{n^2}{2}\right)-8$ when $n$ is even, and to
$\lfloor\nicefrac{n^2}{2}\rfloor-4=\left\lceil\nicefrac{n^2}{2}\right\rceil-5$
when $n$ is odd.
Note that $s$ is decreasing in every outer-loop iteration,
and so $\nicefrac{1}{s}$ is increasing, which implies that
no player receiving a portion valued $\nicefrac{1}{s}$ at a
later outer-loop iteration will envy any of those players that
received a portion in one of the previous outer-loop iterations.
To verify the latter, let us denote by $C^{\prime}_{t}$ the subcake
to be divided and by $s_{t}$ the number of players participating
in outer-loop iteration~$t$ with
$2 \leq t \leq \left\lceil \nicefrac{(n-4)}{2} \right\rceil$.
Considering any outer-loop iteration $t-1$
with cake $C^{\prime}_{t-1}$ being renormalized such that
$v_i(C^{\prime}_{t-1})=1$, $1 \leq i \leq s_{t-1}$,
each of the two players dropping out in iteration~$t-1$ 
receives a portion he or she values exactly
$\nicefrac{1}{s_{t-1}}$ of $C^{\prime}_{t-1}$ and 
which is valued at most 
$\nicefrac{1}{s_{t-1}}$ of $C^{\prime}_{t-1}$ by the 
remaining
players, 
where $v_i(C^{\prime}_{t-1}) \geq \nicefrac{s_{t-1}}{n}$ 
(see Remark~\ref{rem:remarksblock}.\ref{rem:remarksblock-5}).
Consequently, all players $p_j$, $1 \leq j \leq s_t$, participating in
outer-loop iteration~$t$ value subcake $C^{\prime}_{t}$ to be worth
at least $\nicefrac{s_t}{s_{t-1}}$ of $C^{\prime}_{t-1}$, i.e.,
$v_j(C^{\prime}_{t}) \geq (\nicefrac{s_t}{s_{t-1}}) \cdot v_j(C^{\prime}_{t-1})$,
and the two players dropping out
in iteration~$t$ both receive a portion valued
$\nicefrac{1}{s_{t}}$ of subcake $C^{\prime}_{t}$.
Thus, players dropping out in outer-loop iteration~$t$
receive each a portion that is valued
at least~$\nicefrac{1}{s_{t-1}}$ of $C^{\prime}_{t-1}$
according to their measures.

Once the outer loop has
been completed, either four (if $n$ is even) or three
(if $n$ is odd) players are left.  If there are 
four
players left, three more envy-free-relations can be guaranteed by executing
Steps~9.1 through~9.4, as the three remaining players will not envy the
player receiving the fourth-to-last portion,
but no more than three envy-free-relations can be guaranteed,
again by the proof of Lemma~\ref{lem:DGEF-no-evaluations}.
In addition, the three last players entering
Step~10
will not envy each other, because the Selfridge--Conway protocol
always provides an
envy-free division.  Accordingly, six more envy-free-relations are guaranteed.
Summing up, the protocol
in Figure~\ref{algo:n} has a DGEF
of $\left\lceil \nicefrac{n^2}{2} \right\rceil + 1$.~\end{proofs}

\begin{theorem}
\label{thm:n-strategy-proof}
  The proportional cake-cutting protocol in Figure~\ref{algo:n} is
  strategy-proof in the sense of
  Definition~\ref{def:proportional-strategy-proof}.
\end{theorem}

\begin{proofs}
From Step~9.1 on (when at most four players are left),
the protocol has been proven to be strategy-proof in
Theorem~\ref{thm:n4-strategy-proof}.  With respect to Steps~1 through~8,
three different cases need to be considered, each of which applies
analogously to both assigning the leftmost smallest piece and assigning
the rightmost smallest piece.  Note that a cheating attempt on one of the
pieces does not influence the assignment of the other.  The same three
cases apply when there is a cheating attempt on the rightmost smallest
piece and Steps~4.1 through~4.3 need to be executed.  In the following
case distinction,
given any iteration~$t$ of the outer loop and 
cake~$C^{\prime}_{t}=[\lambda, \varrho]$ with
$v_i(C^{\prime}_{t}) = 1$ for all players~$p_i$, $1 \leq i \leq s_t$, we will
consider only the situation when there is exactly one player not telling the
truth with respect to the leftmost smallest piece, trying to get
more than a proportional share.  
Let $p_c$ be this cheating player,
where $1 \leq c \leq s_t$.  

\begin{fall}
\item \label{n-strategy-proof-case1}
  If $p_c$ is the player receiving portion $C_c=[\lambda,\lambda_c]$ with
  $v_c(\lambda,\lambda_c) > \nicefrac{1}{s_t}$ in Step~5 (and dropping
  out in Step~7), then $p_c$ would
  receive more than a proportional share. However, all $s_t-1$
  other players are
  still guaranteed a proportional share, since all of them consider
  portion $C_c$ as being worth at most~$\nicefrac{1}{s_t}$ of $C^{\prime}_{t}$
  according to their measures, so they all consider the remaining
  part of the cake to be worth at least~$\nicefrac{(s_t-1)}{s_t}$ 
  of $C^{\prime}_{t}$.
\item \label{n-strategy-proof-case2}
  If the cheater, $p_c$, would have received portion
  $C_c=[\lambda,\lambda_c]$ in Step~5
  when telling the truth (so $v_c(\lambda,\lambda_c)=\nicefrac{1}{s_t}$ and 
  $v_i(\lambda,\lambda_c) \leq \nicefrac{1}{s_t}$ with $1 \leq i \leq s_t$ and
  $i \neq c$) but now is not (since $p_c$ is
  making a mark at $\lambda_c^\prime$ with 
  $\lambda_c < \lambda_c^\prime$), then the cheater could end up 
  with even less than
  a proportional share.
  Let us see 
  why this holds true.  In this case, another player, $p_j$ with $j \neq c$,
  receives portion
  $C_j=[\lambda,\lambda_j]$, $\lambda_c \leq \lambda_j \leq \lambda_c^\prime$,
  which determines the left boundary of $C^\prime_{t+1}$ 
  (the remaining cake to be continued with) 
  to be $\lambda_j$.
  According to the measure of
  player~$p_c$,
  $[\lambda_j, \varrho]$~is worth at most~$\nicefrac{(s_t-1)}{s_t}$ 
  of $C^{\prime}_{t}$,
  because
  he or she values $[\lambda,\lambda_j]$ 
  to be
  worth at least
  $\nicefrac{1}{s_t}$ of $C^{\prime}_{t}$, as assumed beforehand.

  However, since the protocol works in a parallel way,
  this loss in value may be
  compensated for by some gain in value with respect to the rightmost
  smallest piece $C_k=[\varrho_k, \varrho^{\prime \prime}]$
  that is assigned in Step~6
  (where $\varrho_{k}$ is the left boundary of the rightmost 
  smallest piece marked by some player $p_{k}$, $\|\{c,j,k\}\|=3$,
  and $\varrho^{\prime \prime}$
  is the current right boundary, i.e.,
  either $\varrho^{\prime \prime} = \varrho$
  or $\varrho^{\prime \prime} = \varrho^{\prime}$).
  That is, if 
  $v_c(\lambda_c, \lambda_j) \leq v_c(\varrho_c, \varrho_k)$
  then the loss player $p_c$ experiences by cheating with respect
  to the leftmost smallest
  piece is made up for by an accidentally sufficient gain relative to the 
  assignment of the rightmost smallest piece.\footnote{Note that the
  compensation for
  the cheater's loss
  may also be accumulated 
  over the following rounds as long as the cheater has not been assigned a
  portion yet.}
  However, if
  $v_c(\lambda_c, \lambda_j) > v_c(\varrho_c, \varrho_{k})$ then no 
  sufficient compensation takes place and player $p_c$ considers 
  $C^{\prime}_{t+1}=[\lambda_j, \varrho_{k}]$, which is the part of the cake
  to be continued with and 
  to be divided among $s_{t}-2$~players,
  as being worth at most~$\nicefrac{(s_t-2)}{s_t}$ of $C^{\prime}_{t}$.
  Thus, in this case player $p_c$ is not guaranteed a portion
  that is worth at least $\nicefrac{1}{n}$ of $C$
  in total (according to his or her measure), even though the protocol
  would be proportional if all players were playing by the rules and
  strategies
  required by the protocol.
  Again, all other players are
  still guaranteed a proportional share.  Players~$p_j$ and $p_{k}$
  both value their portion to be
  $\nicefrac{1}{s_t}$ of $C^{\prime}_{t}$, and the 
  $s_t-3$ remaining players consider $C^{\prime}_{t+1}=[\lambda_j, \varrho_{k}]$
  (which is to be divided among $s_t-2$~players)
  as being worth at least $\nicefrac{(s_{t}-2)}{s_t}$ of $C^{\prime}_{t}$.

  Note that if the cheater marks the rightmost smallest piece in the very 
  same outer-loop iteration in which he or she is cheating on the leftmost
  smallest piece and happens to receive the rightmost smallest piece, then
  there is no influence on the portions to be assigned in this
  outer-loop iteration, as
  even the cheater will receive a proportional share being worth 
  $\nicefrac{1}{s_t}$ of $C^{\prime}_{t}$ according to his or her measure.
\item \label{n-strategy-proof-case3}
  If the cheater does not receive the leftmost smallest portion
  $C_j=[\lambda,\lambda_j]$ by cheating 
  nor would have received this portion when telling the truth, 
  then this would not at all affect the leftmost portion to be
  assigned in this particular outer-loop iteration.  
  The player receiving portion
  $C_j=[\lambda,\lambda_j]$ in Step~5
  values this portion as being~$\nicefrac{1}{s_t}$ of $C^{\prime}_{t}$,
  and the remaining players (except for some player $p_k$, $j \neq k$,
  that is assigned the rightmost smallest piece) continue the procedure with
  $C^{\prime}_{t+1}=[\lambda_j,\varrho_k]$ (where $\varrho_k$ is the left boundary of
  the rightmost smallest piece marked by~$p_k$), which each
  of them values at least $\nicefrac{(s_t-2)}{s_t}$ of $C^{\prime}_{t}$,
  even the cheater.
 \end{fall}
This concludes the proof of Theorem~\ref{thm:n-strategy-proof}.~\end{proofs}

Figure~\ref{algo:n:strongfair} shows how to adapt the protocol in
Figure~\ref{algo:n} so as to achieve a strong fair division.  Again,
note that if the inner loop (Steps~4.1 through~4.3) has not been
executed in an outer-loop iteration (Steps~1 through~8), we have the
special case of zero inner-loop iterations and $\varrho =
\varrho^{\prime}$.  In this case, portion
$C_k=[\varrho_k-m_r,\varrho]$ assigned to player $p_k$ in Step~6 
is the same as $C_k=[\varrho_k-m_r,\varrho^{\prime}]$ in the
general case, and the values for $\varrho:=\varrho_k-m_r$ and
$C^{\prime}:=[\lambda_j+m_{\ell}, \varrho_k-m_r]$ that are set in
Step~8 in this case are special cases of
$\varrho:=\varrho-\varrho^{\prime}+\varrho_k-m_r$ and
$C^{\prime}:=[\lambda_j+m_{\ell}, \varrho_k-m_r] \cup
[\varrho^{\prime}, \varrho]$ (again, since $[\varrho^{\prime},
\varrho]$ degenerates to a single point if $\varrho =
\varrho^{\prime}$, which is valued zero by the axiom of divisibility,
see Footnote~\ref{foo:divisibility}).

\begin{theorem}
\label{thm:n-strong-fair}
For $n > 3$ players, modifying the first iteration of the
cake-cutting protocol in Figure~\ref{algo:n} according to
Figure~\ref{algo:n:strongfair} yields a strong fair division, provided
that in this iteration exactly one player
makes a mark that is closest to the left boundary and
exactly one distinct player makes a mark that is closest to
the right boundary.
\end{theorem}

\begin{proofs}
For $n=4$ players, the protocol in Figure~\ref{algo:n} is the same as
the protocol in Figure~\ref{algo:n4} and thus can be modified
so as to yield a strong fair division according to 
Theorem~\ref{thm:n4-strong-fair}, see Figure~\ref{algo:n4:strongfair}.

For $n>4$ players, consider the very first outer-loop iteration (Steps~1
through~8) of the protocol.  Let $p_j$ be the 
unique player whose mark in Step~1 is closest to the left boundary~$\lambda$, 
and let $p_k$, $j \neq k$, be the unique player whose mark in Step~1 is
closest to the right boundary~$\varrho$ (if the inner
loop needs to be executed, let $p_k$ be the unique player whose mark
in Step~4.2 is closest to the current right boundary~$\varrho^\prime$).
According to Step~2 in Figure~\ref{algo:n:strongfair}, let $p_\ell$ be
any player such that $\|[\lambda,\lambda_j]\| < \|[\lambda,\lambda_{\ell}]\|$
and there is no player $p_z$, $1 \leq j,\ell,z \leq s$, $\|\{j,\ell,z\}\|=3$,
with $\|[\lambda,\lambda_z]\| < \|[\lambda,\lambda_{\ell}]\|$,
where ties can be broken arbitrarily. 
Analogously, according to Step~3 (respectively, Step~4.3) in
Figure~\ref{algo:n:strongfair}, let $p_r$ be any player such that
$\|[\varrho_k,\varrho]\| < \|[\varrho_r,\varrho]\|$
(respectively, 
$\|[\varrho_k,\varrho^\prime]\| < \|[\varrho_r,\varrho^\prime]\|$)
and there is no player $p_z$, $1 \leq k,r,z \leq s$, $\|\{k,r,z\}\|=3$,
with $\|[\varrho_z,\varrho]\| < \|[\varrho_r,\varrho]\|$
($\|[\varrho_z,\varrho^\prime]\| < \|[\varrho_r,\varrho^\prime]\|$). 
Steps~5 and~6 in Figure~\ref{algo:n:strongfair} assure that
players~$p_j$ and~$p_k$ each are assigned a portion that,
according to their measures, is worth strictly
more than $\nicefrac{1}{s}$ of $C^{\prime}$ (which is worth at least
as much as $\nicefrac{1}{n}$ of $C$ by the argument of
Remark~\ref{rem:remarksblock}.\ref{rem:remarksblock-5}),\footnote{Recall
that we assumed the axiom of
positivity, which requires nonempty pieces of cake to have a nonzero
value for each player, see Footnote~\ref{foo:positivity} for more
discussion of this point.} 
since $p_j$ and $p_k$ each receive a portion that is bigger and thus
is worth more than the one they have marked as being worth
exactly~$\nicefrac{1}{s}$ of $C^{\prime}$ in Step~1.

By the same argument,
the $n-2$~remaining players continue by applying a proportional protocol
to a part of the cake that all of them consider to be worth strictly more
than~$\nicefrac{(n-2)}{n}$ of $C$. Thus, each of the players receives a portion
that is worth strictly more than $\nicefrac{1}{n}$ of $C$ according to his or
her measure, which results in a strong fair division.~\end{proofs}

\begin{figure}[h!tp]
  \centering
  \begin{tabular}{|lp{118mm}|}
    \hline
    {\bf Step~2.} & Find any players $p_j$ and $p_{\ell}$ such that 
$\|[\lambda,\lambda_j]\| < \|[\lambda,\lambda_{\ell}]\|$ and there is no player
$p_{z}$,$1 \leq j,{\ell},z \leq s$, $\|\{j,{\ell},z\}\|=3$,
with $\|[\lambda,\lambda_{z}]\| < \|[\lambda,\lambda_{\ell}]\|$.
Set $m_{\ell}:=\nicefrac{(\lambda_{\ell} - \lambda_j)}{2}$. \\
    {\bf Step~3.} & Find any players $p_k$ and $p_{r}$ such that
$\|[\varrho_k,\varrho]\| < \|[\varrho_{r},\varrho]\|$ and there is no player
$p_z$, $1 \leq k,r,z \leq s$, $\|\{k,r,z\}\|=3$,
with $\|[\varrho_z,\varrho]\| < \|[\varrho_{r},\varrho]\|$.
If there is more than one player fulfilling this condition for player~$p_k$,
and player $p_j$ is one of them, choose player $p_k$ other than $p_j$.
Set $m_r:=\nicefrac{(\varrho_k - \varrho_{r})}{2}$. \\
    {\bf Step~4.3.} & Find any players $p_k$ and $p_{r}$ such that 
$\|[\varrho_k,\varrho^{\prime}]\| < \|[\varrho_{r},\varrho^{\prime}]\|$ and
there is no player $p_z$, $1 \leq k,r,z \leq s$, $\|\{k,r,z\}\|=3$,
with $\|[\varrho_z,\varrho^{\prime}]\| < \|[\varrho_{r},\varrho^{\prime}]\|$.
If there is more than one player fulfilling this condition for player~$p_k$,
and player $p_j$ is one of them, choose player $p_k$ other than $p_j$. 
Set $m_r:=\nicefrac{(\varrho_k - \varrho_{r})}{2}$.\\
    {\bf Step~5.} & Assign portion $C_j=[\lambda,\lambda_j+m_{\ell}]$ to
player~$p_j$. \\
    {\bf Step~6.} & If $\varrho = \varrho^{\prime}$, assign
portion $C_k=[\varrho_k-m_r,\varrho]$ to player $p_k$, else assign portion
$C_k=[\varrho_k-m_r,\varrho^{\prime}]$ to player $p_k$. \\
    {\bf Step~8.} & If $\varrho = \varrho^{\prime}$, set
$C^{\prime}:=[\lambda_j+m_{\ell}, \varrho_k-m_r]$ and
$\varrho:=\varrho_k-m_r$, else set 
$C^{\prime}:=[\lambda_j+m_{\ell}, \varrho_k-m_r] \cup 
[\varrho^{\prime}, \varrho]$
and $\varrho:=\varrho-\varrho^{\prime}+\varrho_k-m_r$.
Set $\lambda:=\lambda_j+m_{\ell}$, $\varrho^{\prime} := \varrho$, and 
$s:=s-2$. \\
    {\bf Step~9.2.} & Find any players $p_j$ and $p_k$ such that 
$\|[\varrho_j,\varrho]\| < \|[\varrho_k,\varrho]\|$ and there is no
player $p_{\ell}$, $1 \leq j,k,\ell \leq s$, $\|\{j,k,\ell\}\|=3$, with
$\|[\varrho_{\ell},\varrho]\| < \|[\varrho_k,\varrho]\|$. 
Set $m_r:=\nicefrac{(\varrho_j - \varrho_k)}{2}$. \\
    {\bf Step~9.3.} & Assign portion $C_j=[\varrho_j-m_r,\varrho]$ to 
player $p_j$.
Let player $p_j$ drop out. \\
    {\bf Step~9.4.} & Set $\varrho:=\varrho_j-m_r$ and
$C^{\prime}:=[\lambda, \varrho]$. Set $s:=s-1$. \\
    \hline
  \end{tabular}
  \caption{Modified steps in the first iteration of the protocol in
 Figure~\ref{algo:n} to achieve a strong fair division for $n > 3$ players.}
  \label{algo:n:strongfair}
\end{figure}

\section{Proof of Theorem~\ref{thm:survey-DGEF}}
\label{sec:survey}

In this section, we determine the degrees of guaranteed envy-freeness
of the proportional cake-cutting protocols listed in
Table~\ref{tab:DGEF-survey}.  Thus, we prove
Theorem~\ref{thm:survey-DGEF} via Lemmas~\ref{lem:last-diminisher}
through~\ref{lem:divide-and-choose}.
We investigate proportional protocols only, because
proportional cake-cutting protocols have a DGEF of at least
$n$ according to Proposition~\ref{prop:DGEF-Minimum-Maximum} and, thus,
show some degree of fairness
already. 

Over the years, several cake-cutting protocols
have been proven to be proportional and finite bounded for
any number $n$ of players.  
A detailed description of
various finite bounded proportional protocols
can be found in the books by Brams and
Taylor~\cite{bra-tay:b:fair-division} and
Robertson and Webb~\cite{rob-web:b:cake-cutting}.
Our analysis of the protocols in Table~\ref{tab:DGEF-survey}
provides a basis for further algorithmic improvements in terms
of the degree of guaranteed envy-freeness, and the protocol in
Figure~\ref{algo:n} is a first step in this direction.

In the following subsections, we give a brief
description of the protocols listed in
Table~\ref{tab:DGEF-survey} and provide a
detailed analysis of their
{DGEF}.  
Note that the value of cake~$C$ to be divided is normalized such that
$v_i(C)=1$ for all players~$p_i$ with $1 \leq i \leq n$.

\subsection{Last Diminisher}

The protocol works as follows:
The first player cuts a piece he or she
considers being worth exactly $\nicefrac{1}{n}$ in his or her
measure. This piece
is given to the $n-1$ other players, one
after the other.  Now, each player has the choice to either pass the
piece on as it is, or to trim it before passing it on. If a player
considers the piece to be worth more than $\nicefrac{1}{n}$, he or she
trims it to exactly $\nicefrac{1}{n}$ according to his or her
measure.  When the last player has evaluated this piece, it is given to the
player who was the last trimming it, or to the player who cut it in
the first place if no trimmings have been made.  The trimmings are
reassembled with the remainder of the cake, and the procedure is
applied in the same way for the $n-1$ remaining players and the
reassembled remainder of the cake.  This process is repeated until
only two players remain.  In the final round (of $n-1$ rounds in total),
these last two players apply the simple cut-and-choose
protocol to the remainder of the cake.  For guaranteeing each player a
proportional share of the cake,
the order of the players is of no significance.

\begin{lemma}
\label{lem:last-diminisher}
The Last Diminisher protocol has a degree of guaranteed envy-freeness
of $2 + \nicefrac{n(n-1)}{2}$.
\end{lemma}

\begin{proofs}
Concerning the analysis of the degree of guaranteed envy-freeness, it
is quite evident that in the first round $n-1$ envy-free-relations are
guaranteed, since each of the $n-1$ players not receiving the first
piece consider this piece to be of value at most~$\nicefrac{1}{n}$.
Analogously, in
the
$k$th round, $1 < k < n$,
$n-k$ additional envy-free-relations are guaranteed.  The number of
guaranteed envy-free-relations is consecutively decreasing by
one per round, as the players
who already received a piece are not involved in the evaluation
process of subsequent rounds
(see the proof of Lemma~\ref{lem:DGEF-no-evaluations}). 
This sums up to
$\sum_{i=1}^{n-1}{i}=\nicefrac{n(n-1)}{2}$ guaranteed
envy-free-relations.  In addition,
in the final round one more guaranteed envy-free-relation
is created, as the simple cut-and-choose protocol guarantees that both of the
players will not envy each other.
Finally, note that the player receiving the first portion
is not involved in the evaluations of any other portions, but since the
Last Diminisher protocol is proportional, this player too cannot envy
each of the other players according to the argument in the proof of
Proposition~\ref{prop:DGEF-Minimum-Maximum}.  Thus, one more 
guaranteed envy-free-relation must be
added. Consequently, the Last Diminisher protocol guarantees
$2+\nicefrac{n(n-1)}{2}$ envy-free-relations.~\end{proofs}

\subsection{Lone Chooser}

The Lone Chooser protocol was first proposed by
Fink~\cite{fin:j:fair-division} (as cited
in~\cite{saa:b:cake-cutting-protocol,bra-tay:b:fair-division}).
It can be
described as follows. For two players, the protocol is just the simple
cut-and-choose protocol.
For $n > 2$ players, the protocol has $n-1$
rounds.  The first round simply describes the cut-and-choose protocol
executed by players $p_1$ and~$p_2$, which results in two 
pieces, $c_1$ and~$c_2$, with $C = c_1 \cup c_2$. 
Assuming player $p_1$ received 
piece $c_1$
and player $p_2$ received 
piece $c_2$,
in the second round player $p_1$ has to divide piece
$c_1$ with player $p_3$, and player $p_2$ has to divide piece $c_2$
with player $p_3$.
To this end, $p_1$ cuts $c_1$ into three
pieces each of which he or she considers to be worth at least
$\nicefrac{1}{6}$, and so does $p_2$ with~$c_2$. Player $p_3$
then chooses one of the pieces of player $p_1$ and one of the pieces
of player~$p_2$, both being most valuable according to $p_3$'s measure.
This guarantees each of the players $p_1$, $p_2$, and
$p_3$ a portion of at least $\nicefrac{1}{3}$ in their measures. Carrying
on in this way, when the 
final
round has been entered, 
each of the 
players $p_1, p_2, \dots, p_{n-1}$ 
is
in possession of a portion that 
he or she
considers to be worth at
least $\nicefrac{1}{(n-1)}$ in
his or her measure. 
Let us refer to those
$n-1$ players as the ``cutters'' of round~$n-1$. Finally, each of
the cutters $p_1, \dots, p_{n-1}$ cuts his or her portion into $n$
pieces 
each of
value~$\nicefrac{1}{(n^2-n)}$,
and player~$p_n$, the ``chooser'' of round $n-1$, chooses
one piece of highest value (according to his or her measure)
from each plate of the $n-1$ cutters of this round.

\begin{lemma}
\label{lem:lone-chooser}
The Lone Chooser protocol has a degree of guaranteed envy-freeness
of~$n$.
\end{lemma}

\begin{proofs}
In the course of the Lone Chooser protocol, none of the players
evaluate the portion of any of the other players, which determines
the DGEF of the Lone Chooser protocol to be~$n$ as a result of
Lemma~\ref{lem:DGEF-no-evaluations}.~\end{proofs}

\subsection{Lone Divider}

A complete algorithmic description of this protocol would be
rather comprehensive. That is why we will give just a rough sketch of
the procedure. For a more detailed description, the reader is referred
to Kuhn~\cite{kuh:b:games-fair-division}, see also, e.g.,
\cite{bra-tay:b:fair-division,rob-web:b:cake-cutting,daw:j:lone-divider}.

The Lone Divider
protocol works as follows: Some player $p_d$, $1 \leq d \leq n$, is
chosen to be the first-round 
``divider.''  The divider cuts 
cake~$C$
into $n$ portions
that he or she considers each to be worth $\nicefrac{1}{n}$, i.e.,
$v_d(C_j)=\nicefrac{1}{n}$ 
with $1 \leq j \leq n$. Subsequently, all
other players (the first-round  ``choosers'') are
asked to identify any of the $n$ portions they find acceptable, that is,
each player identifies all portions that are worth at least
$\nicefrac{1}{n}$ in this player's
measure.  Obviously, every chooser~$p_c$, $c \neq d$,
needs to accept at least one
portion~$C_j$.  Depending on the players' choices, there are different
ways for how the protocol continues. In the simplest case, the players'
choices allow for a \emph{fully decidable division}, i.e., for a
division such that each chooser $p_c$ receives one
of the portions he or she
previously identified as being acceptable.  Divider
$p_d$ then receives the portion that has not been assigned to any of
the
choosers. All players drop out and the division is complete.
However, if the players' choices do not allow
a fully decidable division, there is either a \emph{partially 
decidable division} (which means that only some---not all---of the
choosers are assigned a portion and drop out), or a
\emph{fully undecidable division}
(which means that there are at least two portions
that have not been identified as being acceptable by any of the choosers
and that none of the choosers receive a portion).
In the case of a partially decidable division, the choosers accomplish
only a partial allocation of the cake, i.e., those choosers
that identified acceptable portions in a nonconflicting way are
assigned a portion they have marked as being
acceptable and drop out.  In addition,
this round's divider is assigned any one of the other portions and
drops out, the remaining portions (that could not be assigned to
players) are reassembled, and a new round with the remaining players, 
among which a new divider is to be chosen, is
started.  In the case of a fully undecidable division, none of the
choosers are assigned a portion; only this round's divider receives one
of the 
two portions
that have not been identified as being acceptable by any of the choosers
and drops out, and a new round 
with the remaining players, among which a new divider is to be chosen,
is started in which the remaining cake is divided.  This
procedure is repeated until the whole cake has been allocated.
Note that in each round at least this round's divider is assigned
a portion and drops out.

It is easy to see that the Lone Divider protocol is finite bounded and
proportional.

\begin{lemma}
\label{lem:lone-divider}
The Lone Divider protocol has a degree of guaranteed envy-freeness
of $2n-2$.
\end{lemma}

\begin{proofs}
The analysis of the degree of guaranteed envy-freeness is done by a
worst-case scenario in terms of the number of existing
envy-free-relations.  We claim that
the 
maximum
number of guaranteed envy-free-relations exists
in the case that every chooser $p_c$ marks $n-1$ portions as being
acceptable in the first round.  In the following,
let us refer to this situation as the
``worst-case scenario.''  In this scenario,
the rules of the protocol imply that a proportional division 
will be achieved in the very
first round (i.e., in this worst case scenario the first round results 
in a fully decidable division), 
and the following envy-free-relations are
guaranteed to exist in this case.
The divider
will not envy any of the choosers, as he or
she considers each of the portions to be $\nicefrac{1}{n}$, resulting in
$n-1$ 
guaranteed envy-free-relations. Furthermore, none of the $n-1$ choosers
will envy the player (be it the divider or any of the other choosers)
that received the portion he or she
considers to be not acceptable, leading to additional $n-1$
guaranteed envy-free-relations.
Since the DGEF is the 
maximum 
number of envy-free-relations that are
guaranteed to exist
in \emph{every} case, the DGEF of the Lone Divider protocol is at
most $2n-2$.
To argue that the scenario given above indeed represents the worst
case for $n \geq 3$ players (and so the DGEF is equal to $2n-2$), we
will consider all possible cases different from the worst-case
scenario.
We will show that in each of these cases the number of existing
envy-free-relations is, in fact, higher than $2n-2$.  This implies
that none of these other cases considered represent a worst-case scenario.

For notational convenience, we will use the term ``case-enforced
envy-free-relation'' to refer to the number of those envy-free-relations
that necessarily must exist in any of these cases,
regardless of which particular valuation functions the players have
(other than what was causing the
respective case to occur).  Recall that the term
``guaranteed envy-free-relation'' is reserved for the number
of envy-free-relations that necessarily exist in the worst case
(and---as we will see---none of the cases below will describe the worst
case).  Thus, the term ``case-enforced envy-free-relation'' is more general 
than and includes the term ``guaranteed envy-free-relation'':
The number of case-enforced envy-free-relations in the worst case
(i.e., the minimum number of case-enforced envy-free-relations,
where the minimum is taken over all possible cases)
is exactly the number of guaranteed envy-free-relations. 

\begin{fall}
\item \label{case1} The first round results in a fully decidable
division.  Consider the following two subcases.

\begin{unterfall}
\item \label{case1-1} The total number of \emph{inacceptable} portions
   is greater than in the worst-case scenario.  Specifically, let us
   look at the situation that exactly one of the choosers considers
   just one more portion as being inacceptable than in the worst-case
   scenario.  This change decrements the total number of acceptable
   portions by one and thus creates one additional case-enforced
   envy-free-relation, since this chooser will not envy the player
   receiving this particular portion.  However, since the number of
   guaranteed
   envy-free-relations of the worst-case scenario persists also in this
   case, increasing the total number of inacceptable portions also
   increases the total number of case-enforced
   envy-free-relations.  Thus, the present case does not describe a
   worst-case scenario.

\item \label{case1-2} The total number of \emph{acceptable} portions
  is greater than in the worst-case scenario.  Specifically, let us
  look at the situation that exactly one of the choosers considers just one
  more portion as being acceptable than in the worst-case scenario.
  This chooser then accepts
  all portions, and thus necessarily considers each of the portions to
  be worth exactly $\nicefrac{1}{n}$.  Accordingly, this chooser does
  not envy any of the other players, resulting in $n-2$ additional
  case-enforced
  envy-free-relations.  In particular, since this chooser still does
  not envy the player he or she did not envy in the worst-case scenario,
  incrementing the total number of acceptable portions by just one
  increases the total number of case-enforced
  envy-free-relations by one (if $n=3$) or even more (if $n>3$).
  Thus, the present case does not describe a worst-case scenario.
\end{unterfall}

\item \label{case2} The first round does not result in a fully
decidable division.  Thus, the protocol runs over more than one round.
  For $n = 3$ players, an additional round would be caused only by a
  fully undecidable division.  For $n > 3$ players, additional rounds
  are caused by either a fully undecidable division or a partially 
  decidable division.
  In every round, at least the divider of this round is
  assigned a portion and drops out.  On the part of the choosers,
  entering a new round creates additional case-enforced
  envy-free-relations the
  number of which depends on the present circumstances.  Simply put,
  every additional round will
  increase the total number of case-enforced
  envy-free-relations compared with the
  worst-case minimum of $2n - 2$.  This justifies why a division
  obtained by the execution of more than one round does not present a
  worst-case scenario.

  To see why running more than one round will
  lead to more than $2n - 2$ case-enforced
  envy-free-relations, let us first, in
  Cases~\ref{case2}.\ref{case2-1} and~\ref{case2}.\ref{case2-2},
  have a closer look at the case-enforced envy-free-relations created in each
  nonfinal round if more than one round is executed in total.  The
  final round will be handled separately in
  Case~\ref{case2}.\ref{case2-3}.  In particular, if there are
  exactly two rounds, the total number of case-enforced
  envy-free-relations created
  is the sum of those explained in the first paragraph of either
  Case~\ref{case2}.\ref{case2-1} or
  Case~\ref{case2}.\ref{case2-2} (which describe the number of case-enforced 
  envy-free-relations created in the first round) and those explained
  in Case~\ref{case2}.\ref{case2-3} (which 
  describes
  the number of case-enforced envy-free-relations created in the final round).

\begin{unterfall}
\item \label{case2-1} In the case of a \emph{fully undecidable
    division} of the cake as the result of the first round, the first-round
    divider receives one of the portions that have not been marked as
    being acceptable by any of the choosers and drops out.
    All $n-1$ choosers enter
    the second round and will not envy the divider of the first round,
    resulting in $n-1$ case-enforced envy-free-relations.
    Moreover, the first-round divider will not
    envy at least one of the $n-1$ other players, since there must be
    at least one player he or she does not envy, which follows from
    the proof of Proposition~\ref{prop:DGEF-Minimum-Maximum}.
    Thus, $n$ case-enforced envy-free-relations
    result from the first round in this case.  

    Note that Proposition~\ref{prop:DGEF-Minimum-Maximum} can be
    applied only to the first-round divider.  Thus, analogously to the
    above argument, for every additional round with $s<n$ players that
    is caused by a fully undecidable division and that is not the
    final round, $s-1$ case-enforced
    envy-free-relations need to be added.  An analysis of the final
    round follows in Case~\ref{case2}.\ref{case2-3}.

\item \label{case2-2} In the case of a \emph{partially 
    decidable division} of
    the cake as the result of the first round, let $K$ denote the set
    of those $k$ players, $1 < k < n-1$, that are in conflict with
    each other concerning the portions they identified as being
    acceptable in this first round. Let $L$ be the set of the
    $\ell$ remaining players, where $1 < \ell < n-1$ and $n = k+\ell$.
    The players in $L$ can divide a part of the cake without any
    conflict, i.e., each of the players in $L$ is assigned one of the
    portions he or she identified as being acceptable.  Note that
    $\ell=1$ would represent Case~\ref{case2}.\ref{case2-1}, and that the
    divider always is one of the players in $L$.  Following the
    protocol, each of the players in $L$ receives a portion and drops
    out while each of the players in $K$ enters the second round.
    Since none of the players in $K$ would accept any of the
    portions the players in $L$ have received, each of the players in
    $K$ does not envy any of the players in~$L$, resulting in $k\ell
    \geq k+\ell = n$ case-enforced
    envy-free-relations.  The players in $L$ (except for the
    first-round divider) that are assigned a portion and drop out in
    the first round are each guaranteed one envy-free-relation due to
    the argument in the proof of
    Proposition~\ref{prop:DGEF-Minimum-Maximum}, summing up to
    $\ell-1$ additional case-enforced
    envy-free-relations.  Moreover, the first-round divider will not
    envy any of the $\ell-1$ other players in~$L$, resulting in
    $\ell-1$ more case-enforced
    envy-free-relations.  Consequently, $2(\ell-1)+k\ell$ case-enforced 
    envy-free-relations result from the first round in this case.

    Now consider any additional round with $s=k^\prime+\ell^\prime$
    players that is caused by a partially 
    decidable division of the cake and
    that is not the final round, where $k^\prime$ players are in
    conflict with each other and $\ell^\prime$ players accomplish a
    partial allocation of the cake.  Analogously to the above argument
    (except that Proposition~\ref{prop:DGEF-Minimum-Maximum} is no
    longer applicable), $k^\prime \ell^\prime$ case-enforced
    envy-free-relations are to be added for the $k^\prime$ players being in
    conflict, and $\ell^\prime-1$ case-enforced envy-free-relations are
    to be added for this round's divider.  An analysis of the final
    round follows in Case~\ref{case2}.\ref{case2-3}.

\item \label{case2-3} Aside from the case-enforced
  envy-free-relations that are
  created by executing a nonfinal round, additional case-enforced
  envy-free-relations are created in the \emph{final round}, in which
  even the last player receives a portion.  
  In each case, the final round is
  characterized by providing a fully decidable division.
  Consequently, considering a final round with $s$~players at least
  $2(s-1)$ case-enforced envy-free-relations are created according to
  Cases~\ref{case1}.\ref{case1-1} and~\ref{case1}.\ref{case1-2}.  
\end{unterfall}
\end{fall}

As a result, in the example of the protocol running a second round,
either $n+2(s-1) = n + 2(n-2)$ with $s=n-1$ (see
Case~\ref{case2}.\ref{case2-1}), or $2(\ell-1)+k\ell+2(s-1) = k\ell
+ 2(n-2) \geq n + 2(n-2)$ with $s=k$ (see
Case~\ref{case2}.\ref{case2-2}) envy-free-relations are case-enforced.
Thus, since $n \geq 3$, a second round
yields at least $n + 2(n-2) > 2n-2$ case-enforced
envy-free-relations in total.  As
indicated above, every additional round increases the number of
case-enforced envy-free-relations even more.

Cases~\ref{case1} and \ref{case2} (and their subcases)
completely characterize all
situations different from the worst case scenario, since their number of
case-enforced envy-free-relations is always greater than $2n-2$, the
number of envy-free-relations guaranteed to exist in the scenario
given in the first paragraph of this proof.  This justifies that this
scenario indeed represents the worst case.  Note
that the protocol does not require any of the choosers to value any of
the portions they marked as being acceptable (i.e., the only
information provided on these portions is that they are considered to
be worth at least $\nicefrac{1}{n}$), and thus, according to the proof
of Lemma~\ref{lem:DGEF-no-evaluations}, the DGEF of the Lone Divider
protocol is $2n-2$.~\end{proofs}

\subsection{Cut Your Own Piece}

The protocol works as follows: 
Every player $p_i$ marks $n$ adjacent pieces each
valued $\nicefrac{1}{n}$ in his or her measure, resulting in $n(n-1)$
marks in total.  Afterwards, a cut is made at between $n-1$ and $2(n-1)$ 
of the existing marks, resulting in at least $n$~adjacent pieces.
Each player then is assigned a portion that contains at least one of
the pieces he or she marked beforehand plus some optional supplement.
There are different strategies for how to make cuts such that the
resulting division is fair.  The number of cuts to be made depends
on the strategy chosen.  Note that there always is at least
one strategy that guarantees each player $p_i$ a portion valued
at least $\nicefrac{1}{n}$ according to his or her measure, 
i.e., $v_i(C_i) \geq \nicefrac{1}{n}$ for all 
$1 \leq i \leq n$~\cite{ste:b:math-snapshots}.
If $n-1$~cuts are made 
(at $n-1$ of the existing marks),
each player's portion consists of exactly one
piece which is worth at least $\nicefrac{1}{n}$ according to his or
her measure.

\begin{lemma}
\label{lem:cut-your-own-piece}
The Cut Your Own Piece protocol has a degree of guaranteed
envy-freeness of $n$ if no strategy is specified, and a degree of
guaranteed envy-freeness of $2n-2$ for the ``left-right strategy''
(which, for convenience, will be explained in the proof).
Moreover, for this protocol no strategy can give a better DGEF than $2n-2$.
\end{lemma}

\begin{proofs}
As mentioned above, there are different strategies
for how to make cuts such that the
resulting division is fair.  If no particular strategy is
given, this protocol has a DGEF of only~$n$ (according to
Lemma~\ref{lem:DGEF-no-evaluations}),
since all players made their marks
independently, and none of the players were asked to give an evaluation
of the marks of any of the other players. Hence, only the minimum of
$n$ envy-free-relations can be guaranteed.
Steinhaus~\cite{ste:b:math-snapshots} did not mention a strategy for how to
achieve a simple fair division; he just mentioned that there always
exists at least one.

However, when we consider a strategy that always assigns the leftmost
(with respect to the interval $[0,1]$) smallest piece to the player
that marked this piece as being of value $\nicefrac{1}{n}$, and the rightmost
smallest piece to the player that marked this piece as being of value
$\nicefrac{1}{n}$, this protocol guarantees at least $2n-2$
envy-free-relations.  We call this strategy the \emph{left-right strategy}.

In more detail, applying the left-right
strategy we assign the
leftmost 
piece to the player that marked the
smallest piece starting at~$0$, and we assign the
rightmost 
piece to the
player that marked the smallest piece finishing at~$1$.  That way
it is guaranteed that the $n-2$ remaining players each consider the part of
the cake between the assigned leftmost piece and the assigned rightmost
piece as being worth at least $\nicefrac{(n-2)}{n}$.  Thus, this
subpart of the cake can be allocated to these $n-2$ players according to
the marks made in the first instance.  Note that if the player that marked
the leftmost smallest piece happens to be the same as the one that
marked the rightmost smallest piece, the
rightmost 
piece is given to the
player that marked the second smallest piece finishing at~$1$.  If
several marks for the leftmost (respectively, for the rightmost) smallest
piece coincide, any mark can be chosen, without loss of generality. In
this example, the terms ``piece'' and ``portion'' can be used
interchangeably, as this 
protocol assigns contiguous portions.

When applying the left-right strategy
as described above, it is guaranteed that the player receiving the
leftmost smallest piece is not envied by any of the $n-1$ remaining
players, since
all of them value this piece at most $\nicefrac{1}{n}$
according to their measures. Analogously, it is guaranteed that the
$n-2$ players in the ``middle''
do not envy
the player receiving the
rightmost smallest piece as they value this piece at
most $\nicefrac{1}{n}$.
Note that if the player that marked the
leftmost smallest piece is the very same as the one that marked the
rightmost smallest piece, this player may envy the player receiving
the
rightmost 
piece, since in this case the
rightmost 
piece to be assigned is just the second
smallest piece finishing at~$1$.  Thus, only $n-2$ envy-free-relations
can be guaranteed with respect to the player receiving the
rightmost 
piece.  However, one more guaranteed envy-free-relation needs to be
added, as the player that marked and is assigned the leftmost smallest
piece cannot envy each of the other players (according to the argument in
the proof of Proposition~\ref{prop:DGEF-Minimum-Maximum}). 
Consequently, $2n-2$ envy-free-relations can be guaranteed in
total.

The DGEF achieved for the Cut Your Own Piece protocol by the
application of the left-right strategy cannot be enhanced by any
other strategy.  The latter is due to the fact that all pieces have
been marked without any mutual evaluations (as already mentioned
above), and that no common boundaries other than
the left border of the leftmost piece and the right border of the
rightmost piece (with respect to the interval $[0,1]$),
which could be used for subsequent comparisons of the guaranteed sizes
and thus values of the pieces marked, are known.~\end{proofs}

\subsection{Divide and Conquer}
\label{sec:divide-and-conquer}

The Divide and Conquer protocol, which was first
presented by Even and Paz~\cite{eve-paz:j:cake-cutting}
(see also, e.g., \cite{rob-web:b:cake-cutting}),
is based on the idea of dividing cake $C$
by simultaneously partitioning disjoint parts of~$C$.  The procedure
slightly differs depending on whether the number of players
is even or odd.

If there is an even number of players, say $n=2k$ for some
integer~$k$, all players but one divide cake $C$ in the ratio
$\nicefrac{k}{k}$ by a single cut, yielding two pieces of equal value
for each of these $n-1$ players.
The noncutter identifies either the piece to the left of the middle cut
(with respect to the interval $[0,1]$), 
or the piece to the right of the middle cut
as being worth at least half of the cake
according to his or her measure,
and then continues dividing this
piece with those $k-1$ cutters whose cuts fall within this
piece. The other piece will be divided among the $k$ remaining
cutters.  
That is, a new round is started in which those two pieces 
of $C$ are
divided among $k$ players each, simultaneously but independently of
each other.

If there is an odd number of players, say $n=2k+1$, all players but one
divide cake $C$ in the ratio $\nicefrac{k}{(k+1)}$
by a single cut.
The noncutter identifies either the piece to the left of the
$k$th cut as being worth at least $\nicefrac{k}{(2k+1)}$, or the piece
to the right of the $k$th cut as being worth at least
$\nicefrac{(k+1)}{(2k+1)}$.  Accordingly, the noncutter continues
dividing either the piece to the left of the $k$th cut with those $k-1$
cutters whose cuts fall within this piece, or the noncutter 
divides the piece to the right of the $k$th cut with those $k$ cutters
whose cuts fall within this piece. In both cases, the other
piece will be divided among all the remaining cutters.

In this way, the procedure is applied recursively until just one player
remains in each subprocedure, i.e., until all the cake has been
allocated to the players. Note that in the case of $n=2$, this is just
the simple cut-and-choose protocol.

Brams, Jones, and Klamler~\cite{bra-jon-kla:j:minimal-envy} present a
finite bounded proportional cake-cutting protocol that is based on
a divide-and-conquer strategy, and focuses on minimizing the number of
players 
the most-envious player
may envy.  The major difference to the protocol
described above lies in the way of splitting the piece of a particular
subprocedure
into two subpieces. While the original Divide and Conquer protocol
suggests to use one of the cuts made by the cutters, the Minimal-Envy
Divide and Conquer protocol suggests to conduct one more cut strictly 
between the cut chosen by the Divide and Conquer protocol and
the very next right neighboring cut (according to the interval $[0,1]$),
and then to use this additional cut for splitting the particular piece
of the cake into two subpieces for the following round if there is any
(or to be assigned if this has been the final round).

\begin{lemma}
\label{lem:divide-conquer}
The Divide and Conquer protocol and the Minimal-Envy Divide and
Conquer protocol both have a degree of guaranteed envy-freeness of
\mbox{$n \cdot \left\lfloor \log n \right\rfloor + 
2n - 2^{\left\lfloor \log n \right\rfloor + 1}$}.
\end{lemma}

\begin{proofs}
The Divide and Conquer protocol is recursively defined.
Put simply, in each subprocedure the given subpart
of the cake is divided into two pieces 
and so are the players into two groups,
the procedure then is applied
recursively again and again to the resulting pieces and related players 
until in each subprocedure just one player remains. 
In each round, every player participating in any of the subprocedures
of this round will not envy at least one of the players continuing
with the corresponding other piece, as this other piece is of no more
value (according to his or her measure) 
than the one he or she is continuing with.
Thus, in each round, for each player involved in this
round, one envy-free-relation is guaranteed to exist.
Note that, for each subprocedure in any round except the final one,
the numbers of players to be continued with
in the resulting two subprocedures of the following round
depend on whether the total number of players involved in
the given subprocedure is even or odd.

From these remarks it follows that the Divide and Conquer protocol's
degree of guaranteed
envy-freeness, call it
$d(n)$ for $n$ players, can be
described by the following recurrence:
\[
\begin{array}{rcll}
d(1) & = & 0,                   & \\
d(n) & = & d(k) + d(k) + 2k     & \text{for $n = 2k$,} \\
d(n) & = & d(k) + d(k+1) + 2k+1 & \text{for $n = 2k+1$.}
\end{array}
\]
Apparently, this recurrence relation can be simplified to:
\begin{eqnarray}
  d(1)& = & 0, \nonumber \\
  d(n) & = & d(\left\lfloor \nicefrac{n}{2} \right\rfloor) + 
             d(\left\lceil \nicefrac{n}{2} \right\rceil) + n 
             \mbox{\quad for $n \geq 2$.}
  \label{eq:rec-solution-divide-conquer}
\end{eqnarray}

The recurrence in~(\ref{eq:rec-solution-divide-conquer}) 
and similar versions are well known
to occur also in other contexts.\footnote{For example,
they also occur in the context of evaluating the number of
comparisons made by various sorting algorithms that are
based on a divide-and-conquer strategy.  In particular,
this recurrence expresses the number of comparisons
done by the standard merge-sort algorithm.}
It is a matter of routine (see, e.g.,
\cite{gra-knu-pat:b:concrete-maths-comp-science})
to solve it (i.e., to bring it into closed form):
\begin{eqnarray}
  \label{eq:solution-divide-conquer} 
  d(n) & = & n \cdot \left\lfloor \log n \right\rfloor + 
           2n - 2^{\left\lfloor \log n \right\rfloor + 1}
             \mbox{\quad for $n \geq 1$.}
\end{eqnarray}

How does Equation~(\ref{eq:solution-divide-conquer}) reflect
the guaranteed number of
envy-free-relations of the Divide and Conquer protocol?
As mentioned before,
this protocol can be considered
as a collection of several subprocedures that altogether yield a
proportional division of the given cake. This collection of subprocedures
can be represented as a balanced
binary tree (called the ``recursion tree''),
because in every subprocedure the given subpart of
the cake is cut into two pieces for which the procedure is
applied recursively again
and again until just one player remains in each resulting subprocedure.
Since the depth of a
balanced binary tree is logarithmic in the number of leaves, 
$\left\lceil \log n \right\rceil$ rounds are performed in total.
If the number $n$~of players is not a power of two, 
every round except for the last one (i.e.,
$\left\lfloor \log n \right\rfloor$ rounds) 
is
represented by a completely 
filled level of the binary tree in terms of the number of players,
since all $n$~players are participating in these
rounds---in different subprocedures though.  Note that
$n$ is a power of two if and only if it holds that 
$\left\lceil \log n \right\rceil = \left\lfloor \log n \right\rfloor$
(i.e., the final round is numbered $\lfloor \log n \rfloor$),
and only in this case, all $n$~players are involved
in each of the rounds, even in the final round.

Recall that, in each round, every participating player will not envy
at least one of the players continuing with the particular other
piece, since he or she considers this piece to be of no more value
than the one he or she is continuing with,
i.e., in each round one guaranteed envy-free-relation is created 
on behalf of each of the participating players.  For this reason,
$n$~envy-free-relations 
are
guaranteed to be created in
each of the first $\left\lfloor \log n \right\rfloor$ rounds.
This is because in subsequent rounds the
particular subparts to be divided
will never get bigger again, and once two players have ended up in
different subprocedures, they will never meet again in the same
subprocedure of any of the
following rounds.  Moreover, once a player has ended up in some group,
he or she will not make future evaluations of pieces of the cake
to be divided among the players in the other group.
Thus, 
those
envy-free-relations that result from 
any
of the 
first $\left\lfloor \log n \right\rfloor$
rounds are guaranteed 
to persist until all the cake has been allocated.
However, it cannot be determined which of the players continuing
with the other piece of the particular subpart is not envied, 
since no evaluations
of the pieces created in other subprocedures are made. 
In accordance with the proof of Lemma~\ref{lem:DGEF-no-evaluations},
the latter also justifies why no more envy-free-relations can be
guaranteed.  Hence, it can just be guaranteed that each
player does not envy at least one of the players continuing with the other
piece.
Summing up, 
as
exactly $n$ guaranteed envy-free-relations are created in 
each of the first $\left\lfloor \log n \right\rfloor$ rounds,
$n \cdot \left\lfloor \log n \right\rfloor$
guaranteed envy-free-relations are created over all rounds,
except for the 
final
round if $n$ is not a power of two.
Note that 
if $n$ is a power of two, the $(\log n)$th round is the final round and
Equation~(\ref{eq:solution-divide-conquer})
simplifies to $d(n) = n \cdot \log n$, 
so in this case we are done.

In contrast, if $n$ is not a power of two then less than $n$~players 
will be involved in the 
final
round (i.e., in the round numbered
$\lceil \log n \rceil = \lfloor \log n \rfloor + 1$),
since in that case there is
at least one subprocedure that involves an odd number of players.
More specifically, in this case the number of players involved in the
final
round can be expressed by the term
$2n - 2^{\left\lfloor \log n \right\rfloor + 1}$, where 
$2^{\left\lfloor \log n \right\rfloor + 1}$ specifies the number of
players that would be involved in the 
final
round if the
binary recursion tree would be a full binary tree, i.e., if all $n$~players
would be involved in the 
final
round.

In order to analyze the 
final
round for $n$ not being a power of two in
detail, let $i$-subprocedure
denote a subprocedure involving exactly $i$ players.
A $3$-subprocedure can occur only in the
second-to-last round, and if it occurs then one of its three 
players cannot be participating in the 
final
round, i.e., 
only two out of three players are proceeding to the 
final
round. 
Since in a balanced binary tree the depth of all leaves differs by
at most one, the second-to-last round can have only either
$2$-subprocedures and/or
$3$-subprocedures, or $4$-subprocedures and/or $3$-subprocedures. 
Regarding a second-to-last round with at least one $3$-subprocedure
and any number of $2$-subprocedures (which can happen only 
if $n$ is not a power of two),
the number of players involved in the 
final
round
will be twice the number of $3$-subprocedures occurring in the
second-to-last round.
Regarding a second-to-last round with at least one $3$-subprocedure and
any number of $4$-subprocedures (which again can happen only 
if $n$ is not a power of two),
the number of players involved in the 
final
round will be
twice the number of $3$-subprocedures plus four times
the number of $4$-subprocedures.
Consequently, 
the number of $3$-subprocedures and the number of
$4$-subprocedures in the second-to-last round
determine how many players are participating in the 
final
round, and thus also determine the number of 
guaranteed
envy-free-relations 
to be created
in 
the
final
round.

If $n$ is not a power of two,
analogously to the argument for the 
first $\left\lfloor \log n \right\rfloor$ rounds, also in the 
final
round one guaranteed envy-free-relation 
is created with respect to
each participating player.
Thus, for any number $n \geq 1$ of
players, at least $2n - 2^{\left\lfloor \log n \right\rfloor + 1}$
guaranteed envy-free-relations are 
created 
in the
$(\lfloor \log n \rfloor + 1)$th round.\footnote{Note 
  that $n$ is a power of two if and only if
  $2n - 2^{\left\lfloor \log n \right\rfloor + 1} = 0$,
  and that in this case 
  there is no $(\lfloor \log n \rfloor + 1)$th round.}
According to the proof of Lemma~\ref{lem:DGEF-no-evaluations},
no more than that many envy-free-relations can be guaranteed.
Altogether, this sums up to $n \cdot \left\lfloor \log n \right\rfloor +
2n - 2^{\left\lfloor \log n \right\rfloor + 1}$
guaranteed envy-free-relations in total, which is the number stated
in Equation~(\ref{eq:solution-divide-conquer}).

The degree of guaranteed envy-freeness of the Minimal-Envy Divide
and Conquer protocol can be shown just as for the original
Divide and Conquer protocol.  The difference in the way of splitting
the particular piece of the cake into two subpieces does not affect
the number of guaranteed envy-free-relations.  Consequently, although
the  Minimal-Envy Divide and Conquer protocol does decrease envy
according to the definition of Brams, Jones, and
Klamler~\cite{bra-jon-kla:j:minimal-envy}
(see Section~\ref{sec:discussion} for more discussion of this point),
its DGEF is
$n \cdot \left\lfloor \log n \right\rfloor +
2n - 2^{\left\lfloor \log n \right\rfloor + 1}$,
just as for the original Divide and Conquer protocol.~\end{proofs}

\subsection{Recursive Divide and Choose}

This protocol has been presented by 
Tasn\'{a}di~\cite{tas:j:proportional-protocol} and
describes a recursive procedure for how to always achieve a proportional
division. It works as follows:
In the case of $n=2$, this is just the simple cut-and-choose protocol.
In the case of $n=3$, one of the players, the ``divider,''
divides the cake into three equal pieces according to his or her
measure, and each of the two other players, the ``choosers,'' marks two
pieces he or she considers to be worth the most, where ties may be broken
arbitrarily. If both choosers
marked the same two pieces, they divide these by applying the
simple cut-and-choose protocol, and the divider receives the remaining
piece. If the choosers marked different pieces, they
divide the
piece they both have marked via the simple cut-and-choose protocol,
and each of the choosers
divides the piece marked by just
him- or herself
with the divider, again via applying the simple cut-and-choose
protocol.

In the case of $n > 3$ players, this procedure is repeated
recursively until all comes down to the simple cut-and-choose
protocol involving two players only. In more detail, 
in the first round the divider cuts
the cake into $n$ equal pieces according to his or her measure, and
each of the $n-1$ choosers marks $n-1$~pieces he or she
considers to be worth the most, where ties may be broken arbitrarily.
Afterwards, a new round is started and each of the $n$~pieces 
is divided among those $n-1$ 
choosers
that identified this piece
as being acceptable.
Concerning pieces that have been marked by less than $n-1$~choosers,
the divider fills out these empty slots by an appropriate number of
clones.
In other words, each of the $n$~pieces enters the next round of the
protocol and induces a new subprocedure in the scope of which
the particular piece is being divided among $n-1$ players.  
All $n$~subprocedures are executed simultaneously but
independently of each other.
Note that if in the very first round
all $n-1$~choosers marked the same $n-1$~pieces as being
acceptable than there is exactly one piece that has not been
marked by any of the choosers.  In this case, this piece is directly
assigned to the divider and the divider drops out, 
whereas all choosers enter the next round for dividing the 
$n-1$~remaining pieces among them. 

Analogously, in the $k$th round, $1 < k < n$, there are 
$\prod_{i=2}^{k}{(n-i+2)}$~subprocedures, i.e.,
$\prod_{i=2}^{k}{(n-i+2)}$~pieces are to be divided simultaneously
but independently among $n-k+1$~players each.
In every subprocedure any one player is determined to be the divider 
and cuts the particular piece of this subprocedure into
$n-k+1$~equal subpieces according to his or her measure.
Afterwards, each of the $n-k$~choosers of this particular subprocedure 
marks $n-k$~pieces he or she considers to be worth the most, 
where ties may be broken arbitrarily.
Each of the $n-k+1$~pieces will induce a new subprocedure in the 
next round and will be divided among those $n-k$~players that 
marked this piece as being acceptable---where the divider fills out
all empty slots regarding pieces that have been marked by less
than $n-k$~choosers.
Again, if in some round $k$, $1 < k < n$, in any of the subprocedures,
all $n-k$~choosers agree on the same $n-k$~pieces, 
then there will be exactly one piece that has not been marked
by any of the choosers.  In this case, the unmarked piece is directly
assigned to the divider of this particular subprocedure, 
and in the following rounds, the divider will not be involved
in any of the subprocedures that results from this one,
i.e., only $n-k$~pieces enter the next round, in which these are
to be divided among the $n-k$~choosers of the previous round.
Nevertheless, this divider will enter the next round and will
participate in all those subprocedures that result from procedures
he has not 
dropped out from yet.

Applying this procedure recursively until only two players remain
in each subprocedure (i.e., running~$n-1$ rounds), and dividing the 
corresponding piece of the cake between these two via the simple 
cut-and-choose protocol, the Recursive Divide and Choose protocol 
provides a proportional division of the cake.

\begin{lemma}
\label{lem:divide-and-choose}
The Recursive Divide and Choose protocol has a degree of guaranteed
envy-freeness of~$n$.
\end{lemma}

\begin{proofs}
For $n \geq 3$~players, no evaluations of entire portions are
made---except for the one special case when the first-round divider
drops out in the very first round (which, for the sake of
self-containment, will be considered separately below)---and 
thus the scenario from the proof of Lemma~\ref{lem:DGEF-no-evaluations} 
is applicable.  Simply put, only $n$~envy-free-relations can be 
guaranteed in every case due to the argument in the proof of
Proposition~\ref{prop:DGEF-Minimum-Maximum}, as the missing evaluations
of entire portions allow for any valuation functions, for example, those 
described in the proof of Lemma~\ref{lem:DGEF-no-evaluations}.
If in the very first round all $n-1$~choosers agree on the very same
$n-1$~pieces then the first-round divider will drop out with a portion
that he or she values exactly $\nicefrac{1}{n}$ and that all other
players value at most $\nicefrac{1}{n}$.  This results in $n$ guaranteed
envy-free-relations, since none of the $n-1$~choosers will envy the
first-round divider and the first-round divider will not envy at least
one of the other players by the argument in the proof of
Proposition~\ref{prop:DGEF-Minimum-Maximum}.  However, in this case
no more envy-free-relations can be guaranteed in the following rounds
due to the argument given above.
Consequently, the DGEF of the Recursive Divide and Choose protocol
is~$n$.~\end{proofs}

\section{Related Work and Discussion}
\label{sec:discussion}

The analysis of envy-relations dates back at least to Feldman and
Kirman~\cite{fel-kir:j:fairness-and-envy}.  In contrast to our approach,
they consider the number of envy-pairs in already existing divisions with
the intention of maximizing fairness
afterwards via trading.
In particular, they
do not consider the \emph{design} of cake-cutting protocols that
maximize fairness. In the majority of cases, research
in the area of cake-cutting from an economic perspective
is concerned more with the existence of certain divisions and
their properties than with how to achieve these divisions.

A different approach measures the intensity of envy
in terms of the distance between envied portions~\cite{cha:j:measure-envy}.

More recently, Brams, Jones, and
Klamler~\cite{bra-jon-kla:j:minimal-envy} proposed to minimize envy in
terms of the maximum number of players that a player may
envy.
  Their notion of 
  measuring envy differs from our notion of 
  DGEF in various ways, 
  the most fundamental of which is that their notion
  takes an ``egalitarian'' approach to
  reducing the number of envy-relations (namely, via minimizing the
  most-envious player's envy, in terms of decreasing the number of
  this single player's envy-relations).
  In contrast, the DGEF aims at a ``utilitarian''
  approach (namely, via minimizing overall envy,
  in terms of increasing the total number of guaranteed
  envy-free-relations among all players).
  That is to say that,
  although these notions may seem to be very similar at first glance,
  the approach presented
  in~\cite{bra-jon-kla:j:minimal-envy} is not sensitive to 
  a reduction in the number of envy-relations on the part of
  any other than the most-envious player, whereas the DGEF
  does take each single improvement into account
  and adapts accordingly.
  The DGEF, thus, is a more specific, more fine-tuned measure.
  Note also that Brams, Jones, and
  Klamler~\cite{bra-jon-kla:j:minimal-envy}
  focus primarily on presenting a new protocol 
  and less so 
  on introducing a new
  notion for measuring envy.

Another approach is due to Chevaleyre et
al.~\cite{che-end-est-mau:c:envy-free-states}, who
define various metrics for the evaluation of envy in order to
classify ``the degree of envy in a society,'' and they use the term
``degree of envy'' in the quite different setting of multiagent
allocation of \emph{indivisible} resources.

Besides, we stress that our approach of approximating envy-freeness
differs from other lines of research that
also deal with approximating fairness.
For example, Lipton et al.~\cite{lip:c:approximately-fair} propose
to seek for minimum-envy allocations of \emph{indivisible}
goods in terms of the value difference of the utility
functions of envied players,
and Edmonds and Pruhs~\cite{edm-pru:c:not-a-piece-of-cake,edm-pru:c:balanced-allocations-of-cake}
approximate fairness in cake-cutting protocols
by allowing merely approximately fair pieces 
(in terms of their value to the players) and by using only approximate
cut queries (in terms of exactness).  

It may be tempting to seek to decrease envy (and thus to increase
the DGEF) via trading, 
aiming 
to get rid of 
potential circular envy-relations.
Although we do not consider trading to be an integral part of a
cake-cutting protocol, let us for a moment digress to briefly discuss
how trading may potentially affect the number of guaranteed
envy-free-relations.\footnote{To be specific here,
  all occurrences of ``guaranteed
  envy-free-relations'' in this and the next paragraph refer to those
  envy-free-relations that are guaranteed to exist after executing
  some cake-cutting protocol \emph{and in addition, subsequently,
    performing trades that are guaranteed to be feasible}.  This is in
  contrast with what we mean by this term anywhere else in the paper;
  ``guaranteed envy-free-relations'' usually refers to those
  envy-free-relations that are guaranteed to exist after executing the
  protocol only.}
Indeed, if the DGEF is \emph{lower than~$\nicefrac{n(n-1)}{2}$},
the number of guaranteed envy-free-relations can be improved 
to this lower bound, or to an even higher number, by 
resolving circular envy-relations (of which two-way envy-relations are
a special case) by means of circular trades after the execution of the
protocol.
Thus, in this case, involving subsequent trading actions adds on the
number of guaranteed envy-free-relations.
Furthermore, having exactly $\nicefrac{n(n-1)}{2}$ guaranteed
envy-free-relations after all
circular envy-relations have been resolved, three more guaranteed
envy-free-relations
can be gained by applying an envy-free protocol (e.g., the Selfridge--Conway
protocol)
to the three most envied players, which yields to an overall lower bound
of~$3+\nicefrac{n(n-1)}{2}$
guaranteed envy-free-relations.
To give an example for an even higher impact of trading,
when circular trades indeed are involved after
executing either the Divide and Conquer protocol or the Minimal-Envy
Divide and Conquer protocol, their
numbers of guaranteed envy-free relations can be improved
to~$\left(\nicefrac{n(n+1)}{2}\right)-1$, which follows
from~\cite{bra-jon-kla:j:minimal-envy}.

On the other hand,
if the DGEF of a 
proportional cake-cutting protocol is 
\emph{$\nicefrac{n(n-1)}{2}$ or higher}
(such as the DGEF of the protocol 
presented in Figure~\ref{algo:n})
then---depending on the protocol---circular envy-relations may not be
\emph{guaranteed}
to exist, and
if such cycles are not guaranteed to exist,
trading has no impact on the number of
guaranteed envy-free-relations.

However, as mentioned above, we consider trading not to be part of a
cake-cutting protocol, though it might be useful in certain cases (for
example, Brams and Taylor~\cite[page~44]{bra-tay:b:fair-division}
mention that trading might be used ``to
obtain better allocations; however, this is not a procedure but an
informal adjustment
mechanism'').
In particular,
the notion of DGEF refers to (proportional) cake-cutting protocols
without additional trading, i.e.,
the DGEF is defined to make a statement on the performance
of a particular protocol and not about all sorts of actions
to be undertaken afterwards.

Although the well-known protocols listed in Table~\ref{tab:DGEF-survey}
have not been developed with a focus on maximizing the DGEF,\footnote{Quite
remarkably, without any trading actions 
and without involving, e.g., the Selfridge--Conway protocol 
the Last Diminisher protocol
achieves with its DGEF almost (being off only by
one) the trading- and Selfridge--Conway-related
bound of $3+\nicefrac{n(n-1)}{2}$ mentioned
above.}
linking their degrees of guaranteed envy-freeness to the lower bound
provided by involving, e.g., the
Selfridge--Conway protocol and
guaranteed trading opportunities
indicates that the development of cake-cutting protocols with
a considerably higher DGEF or even with a DGEF close to
the maximum of~$n(n-1)$ poses a true challenge.  That is why we feel that
the enhanced DGEF of the protocol presented in Figure~\ref{algo:n}
constitutes a significant improvement.

\section{Conclusions}
\label{sec:conclusion}

Although different disciplines are engaged in the development of
fair cake-cutting protocols for decades now, finite bounded
protocols that 
guarantee an envy-free division for $n>3$ players are still a
mystery. However, finite bounded protocols are the ones we are
looking for in terms of practical implementations. That is why we
in this paper 
have proposed to 
weaken
the requirement of envy-freeness (as much
as needed and as little as possible), while insisting on
finite boundedness. To this end, we introduced the notion of
degree of guaranteed envy-freeness for proportional cake-cutting
protocols. Based on this definition, we
gave a survey of the DGEF in existing finite bounded proportional
cake-cutting protocols, which shows that when one is trying to
approximate the ideal of envy-freeness via the DGEF, 
there is quite a bit room for
improvements.  In particular, we expect that the concept of DGEF 
is suitable to extend the
scope for the development of new finite bounded cake-cutting protocols
by allowing to approximate envy-freeness step by step. In this
context, we proposed a new finite bounded 
proportional cake-cutting protocol, explicitly
demonstrated for
$n=4$ and for arbitrary~$n \geq 3$, which provides a significantly
enhanced degree of guaranteed envy-freeness,
compared with the status quo given by
the survey in Table~\ref{tab:DGEF-survey}
(see also Section~\ref{sec:survey}).  In particular, our protocol
has $\left\lceil \nicefrac{n}{2} \right\rceil - 1$ more guaranteed
envy-free-relations than the Last Diminisher protocol, which previously
was the best finite bounded proportional cake-cutting protocol with respect to
the {DGEF}.
To achieve this significantly enhanced DGEF,
our protocol makes use of parallelization with respect to the
leftmost and the rightmost pieces.
In this regard, adjusting the values of the pieces
to be marked from $\nicefrac{1}{n}$ to $\nicefrac{1}{s}$ (with $s$
players still in the game) and applying an
appropriate inner-loop procedure is crucial to
make the parallelization work.
In addition to an enhanced DGEF, our protocol still has the other useful
properties the Last Diminisher protocol is known to possess, 
such as strategy-proofness.

In general, we suggest to
target improvements on the
``level of envy-freeness'' already in the design of
cake-cutting protocols rather
than trying to improve on the number of 
envy-free-relations 
afterwards for
the division obtained (e.g., via 
trading~\cite{fel-kir:j:fairness-and-envy,cha:j:measure-envy}).
In terms of future research, this approach
encourages to develop new protocols with even higher degrees of
guaranteed envy-freeness---one may even think of modifications that
focus on the development of protocols with ``balanced
envy-freeness'' while
keeping the DGEF high.

\subsubsection*{Acknowledgments}

We are grateful to Ariel Procaccia for many interesting discussions
and pointers to the literature.
In particular, the second author thanks him for drawing his attention
to the fascinating area of cake-cutting during a visit to Hebrew
University of Jerusalem, and he thanks Jeff Rosenschein for hosting
this visit.  We also thank Magnus Roos for helpful comments.

{\small

\bibliographystyle{alpha} 

} %

\end{document}